\documentclass[%
 reprint,
 amsmath,amssymb,
 aps,
]{revtex4-2}

\usepackage{graphicx}
\usepackage{dcolumn}
\usepackage{bm}
\usepackage{amsmath}
\usepackage[colorlinks,
            linkcolor=red,
            anchorcolor=blue,
            citecolor=green
            ]{hyperref}

\begin{document}

\preprint{APS/123-QED}

\title{Enhancing Taiji's Parameter Estimation under Non-Stationarity: a Time-Frequency Domain Framework for Galactic Binaries and Instrumental Noises}

\author{Minghui Du$^1$}
\email{duminghui@imech.ac.cn}%
\author{Ziren Luo$^{1,2,3}$}
\author{Peng Xu$^{1,2,3,4}$}
\email{xupeng@imech.ac.cn}

\affiliation{$^1$Center for Gravitational Wave Experiment, National Microgravity Laboratory, Institute of Mechanics, Chinese Academy of Sciences, Beijing 100190, China}
\affiliation{$^2$Taiji Laboratory for Gravitational Wave Universe (Beijing/Hangzhou), University of Chinese Academy of Sciences (UCAS), Beijing 100049, China}
\affiliation{$^3$Key Laboratory of Gravitational Wave Precision Measurement of Zhejiang Province, Hangzhou Institute for Advanced Study, UCAS, Hangzhou,  310024, China}
\affiliation{$^4$Lanzhou Center of Theoretical Physics, Lanzhou University, Lanzhou 730000, China}

\date{\today}

\begin{abstract}
The data analysis of space-based gravitational wave detectors like Taiji faces significant challenges from non-stationary noise, which compromises the efficacy of traditional frequency-domain analysis. 
This work proposes a unified framework based on short-time Fourier transform (STFT) to enhance parameter estimation of Galactic binary and characterization of instrumental noise  under non-stationarity. 
Segmenting  data into locally stationary intervals, we derive  STFT-based models for signals and noises, and implement Bayesian inference via the extended Whittle likelihood.   
Validated through the analysis of verification Galactic binaries and instrumental noises, our STFT approach outperforms frequency-domain methods by reducing the uncertainty and bias of estimation,  successfully recovering low signal-to-noise ratio signals missed by frequency-domain analysis,  and mitigating  the degeneracy among noise parameters. 
The framework’s robustness against noise drifts and computational efficiency  highlight its potential for integration into future global analysis pipelines.
\end{abstract}

\maketitle


\section{\label{sec:introduction}introduction}
In the upcoming decade, space-based gravitational wave (GW) detectors, including the Laser Interferometer Space Antenna (LISA)~\cite{AmaroSeoane2017LISA,Baker:2019nia}, Taiji~\cite{taiji_0,taiji_1,taiji_2} and Tianqin~\cite{TianQin1},  will expand our exploration of  GW universe to the 0.1 mHz - Hz band by  detecting   a rich spectrum of sources, such as massive black hole binaries (MBHBs)~\cite{mbhb1,mbhb2,mbhb3}, Galactic binaries (GBs, mostly double white dwarfs)~\cite{lisa_gb1,lisa_gb2,lisa_gb3_observation_driven,lisa_gb4}, extreme mass-ratio inspirals (EMRIs)~\cite{emri_rate}, stellar-mass black hole binaries (sBHBs)~\cite{sbhb_rate1,sbhb_rate2}, as well as  stochastic GW backgrounds (SGWBs) of  astrophysical and  cosmological origins~\cite{astrophys_sgwb,cosmological_sgwb}.  
Deriving scientific implications from these detections hinges critically on advanced data analysis methods. 
In this paper, we intend to address one of  the key challenges faced by Taiji's data analysis~\cite{TDCII}, that is the problem caused by  non-stationary noise. 
In principle, the proposed methodology also applies to LISA and Tianqin due to similar mission concepts. 

Different from current observations of LIGO-Virgo-KAGRA,  
the majority of  detectable signals in the space-based band  are continuous GWs.
For instance, due to their quasi-monochromatic nature,   GBs' observational timescales  are expected to be   years.
Observing GBs via GWs offers distinct advantages, as GWs are almost ``transparent'' to interstellar gas, dust, and other stars in the Galaxy~\cite{Breivik:2019oar}. 
Therefore, inferring the properties of GBs provides  ideal probes for the   structure and evolution of the Milky Way~\cite{PhysRevD.111.044023}. 
Another key motivation for precise  estimation of GB parameters stems from the signal-dominated characteristics of Taiji's data. 
With approximately $\mathcal{O}(10^4)$ resolvable and $\mathcal{O}(10^7)$ unresolved GB signals  overlapping  in the data stream~\cite{taiji_confusion_noise,gb_iterative_subtraction}, it is necessary to perform  so-called  ``global fit'' analysis to simultaneously model all the signals and noises components~\cite{gb_globalfit,global_fit_Littenberg:2023xpl,global_fit_Katz:2024oqg,global_fit_geemoo}. 
In other words, inaccurate estimation and subtraction of  single signal could potentially compromise the subsequent analysis.

On the other hand, characterizing instrumental noises also typically relies  on accumulating  long-duration data~\cite{gang_wang_noise_char_1_PhysRevD.106.044054,unequal_arm_unequal_amplitude_PhysRevD.107.123531}.
Accurate  knowledge of  instrumental noises is  essential for  both the statistical inference of  resolvable signals~\cite{matched_filter_Davis1989ARO,matched_filter_Jaranowski_Krolak_2009} and the detection of SGWB~\cite{noise_variation_in_frequency}. 
Besides,  it  also serves as an important indicator for the detector's in-orbit performance. 
Accurate and timely evaluation of noises would enable rapid responses to any potential changes by the mission's ground segment.
Under the signal-dominated scenario, 
joint estimation of noise  and SGWB parameters has been demonstrated in previous studies (\emph{e.g.} Refs.~\cite{gang_wang_noise_char_1_PhysRevD.106.044054,unequal_arm_unequal_amplitude_PhysRevD.107.123531,SGWBinner_Caprini_2019,SGWBinner_Flauger:2020qyi,noise_SGWB_separation_PhysRevD.103.103529,unknown_noise_frequency_shape_Baghi_2023}), while a comprehensive  global analysis of resolvable signals, SGWBs and instrumental noises still remains to be established.  
A a way around this difficulty is to utilize the  ``null'' time-delay interferometry (TDI) channels.
To mitigate the overwhelming laser frequency noise, data transmitted to  Earth  must undergo  TDI processing prior to the scientific analysis~\cite{TDI_Tinto}. 
Among all the divergent schemes of TDI combinations,  ``null'' channels, such as the Michelson-$T$ and Sagnac $\zeta$  channels~\cite{null_channels_PhysRevD.105.023009,null_channels_PhysRevD.105.062006,null_channels_PhysRevD.107.082004,WANG2024107481} are particular valuable since they are insensitive to GWs, offering an direct approach to  understand instrumental noises.

When extracting  information from  long-duration data, the statistical properties of noise inevitably change over time,  posing challenges for data analysis. 
The non-stationarity can be attributed to multiple mechanisms~\footnote{The focus of this paper primarily lies in slowly drifting noises. Transient noises such as glitches can be addressed with alternative methods like fitting and subtraction~\cite{glitch_detection_baghi,glitch_fit_subtract_muratore}.}, including  the cyclostationary  astrophysical foreground caused by the  anisotropic distribution of unresolvable GBs~\cite{PhysRevD.89.022001,Lin:2022huh, cyclostationary_foreground}, 
the variations in payloads, spacecraft (S/C) platforms and in-orbit environments (as demonstrated by LISA Pathfinder~\cite{Armano2016SubFemtoG,Armano2018Calibrating,Armano2018Beyond}), \emph{etc.} 
Additionally,  the noise transfer functions of certain TDI combinations, especially the Michelson-$T$ channel used for noise characterization~\cite{unequal_arm_lisa_taiji,hybrid_relay_noise_characterization}, are sensitive to the variation of  armlengths (which are used as the ``delay''s),  introducing further non-stationarity to  the resulting TDI data. 
Naturally, these effects necessitate regular characterization and updating  of the noise models, 
or otherwise they will lead to bias in parameter estimation or incorrect confidence intervals. 
More critically, non-stationarity would undermine the efficacy of conventional  frequency-domain analysis methods. 
Non-stationarity  violates the fundamental  assumptions of  Whittle likelihood framework~\cite{whittle_likelihood}, as the frequency-domain noise covariance matrix becomes  non-diagonal~\cite{gap_baghi,cyclostationary_foreground}. 
Consequently,  calculating the likelihood function would require   accounting for  the off-diagonal elements, 
substantially increasing computational costs of  Bayesian inference.

A theoretically viable approach to addressing these problems 
is to transition the whole statistical analysis framework to  the time-frequency domain.
After Ref.~\cite{time_frequency_analysis_cornish} provided a concise and systematic overview  of time-frequency GW data  analysis based on the Wilson–Daubechies–Meyer (WDM) wavelet, 
Ref.~\cite{Digman:2022jmp} applied this framework to the characterization of   time-varying GB foreground, and re-evaluated the delectabilities of LISA to GBs and MBHBs. 
Another application to the estimation of LISA-band sBHB was presented in Ref.~\cite{sbhb_pe_wdm}, showing that this framework allows for transient glitches in the data. 
Ref.~\cite{sft_gw_analysis} proposed another approach to deal with noise non-stationarity and data gap, which was 
built on $\mathcal{F}$-statistics and  short-time Fourier transform (STFT).
As examples, the paper assessed its future applicability in the early warning of  binary neutron stars (BNSs)  (for third-generation ground-based detectors) and MBHB (for LISA) in terms of a ``relative error'' metric.  
Also for third-generation ground-based detectors, 
Ref.~\cite{Kumar:2022tto} investigated the impact of non-stationary noise  on the accuracy of BNS parameter estimation, and proposed a robust treatment by normalizing noise power spectral density  (PSD)  along the signal's time-frequency track. 
In a broader sense, time-frequency representations are  widely adopted in template-free approaches such as coherent wave burst~\cite{2008CQGra..25k4029K}. 

Given the potential advantages of time-frequency analysis in space-based detections, and considering its applications in other long-duration data analysis tasks  remain to be fully explored, 
in this paper, we develop a  STFT-based Bayesian analysis framework  to enhance Taiji's (also LISA/Tianqin's) estimation of GB and instrumental noise parameters in the presence of non-stationarity. 
Specifically,  a STFT template for the TDI response of GB signal, and a time-varying noise spectral  model are derived. 
Further, under the   local stationarity condition, we extend  the key statistics such as inner product and  Whittle likelihood  to the time-frequency domain.
The computation of aforementioned models and likelihood can be simply accelerated on GPUs using \texttt{CUPY}, highlighting their potential to be integrated into future  global fit analyses.
When tested on the  estimation of verification GB (VGB)~\cite{Kupfer:2023nqx} parameters,  
our STFT approach achieves  higher accuracy and reduced bias compared to the conventional frequency-domain method. 
For the estimation of noise parameters, we demonstrate our framework on the time-varying $T$ channel (driven by  armlength variation) as an example. 
The results show that 
time-frequency analysis not only establishes a more rigorous theoretical foundation, but also relieves the  degeneracies  among parameters through explicit modeling of temporal features.  
It should be noted that 
this idea has broader applicability: one may leverage the modeled temporal features  to  distinguish  anisotropic stochastic signals from noises, as demonstrated in  Ref.~\cite{Criswell:2024hfn}.

The outline of this paper is as follows: 
Section~\ref{sec:tfdomain} establishes the theoretical formalism of our STFT-based framework,  including the models of GB signals and instrumental noises, as well as the extended Bayesian statistics. 
Section~\ref{sec:gb_estimation} validates the framework on  VGB  parameter estimation, and  demonstrate its advantage over conventional frequency-domain analysis under noise non-stationarity. 
Another data analysis task is illustrated in Section~\ref{sec:noise_characterization}, where we  apply the STFT framework to instrumental noise characterization via the Michelson $T$ channel, showing how explicit modeling of time-dependence mitigates parameter degeneracies.
Conclusions and outlooks for future works are provided in Section~\ref{sec:conclusion}. 
Details of the window function and the full posterior distributions of representative VGBs are presented in Appendix~\ref{appendix:tukey_window} and Appendix~\ref{appendix:posterior}, respectively.

Throughout this paper, we assume all the  data are perfectly synchronized to Barycentric Coordinate Time (TCB), and the laser noise, clock noise, S/C jitter, and tilt-to-length noise have been  sufficiently mitigated during or after TDI processing,  hence excluded from our study.
The time spans of all simulations are set to 1 year.

\section{\label{sec:tfdomain}STFT framework for GW signal estimation and noise characterization}

\subsection{\label{subsec:signal_noise_model}The time-domain  signal and noise models of Taiji}
Despite the extensive literature describing Taiji and LISA's  GW responses and   noise budgets, a systematic introduction  would still be necessary for subsequent analyses. 
We start from the ``single-arm'' Doppler measurement denoted as $\eta_{ij}(t)$, which is essentially the  combination of several  raw  interferometor readouts~\cite{Otto_thesis,Hartwig_thesis,LISA_instrument,Pchannel}. 
Generally, $\eta_{ij}$  
 is the superposition of single-arm GW responses $y_{ij}(t)$ and instrumental noises $n_{ij}(t)$:  
\begin{equation}\label{eq:single_arm_data}
    \eta_{ij}(t) = y_{ij}(t) + n_{ij}(t),  
\end{equation}
where $ij \in \mathcal{I}_2 \equiv \{12, 23, 31, 21, 32, 13\}$, $i$ and $j$ specifying the index of S/C receiving and emitting lasers, respectively. 
$y_{ij}(t)$ can be formulated as~\cite{orbit_approximation,LDC_Radler_manual_v2,TDCI}
\begin{eqnarray}\label{eq:time_domain_single_arm_response}
    y_{ij}(t) &\equiv& \frac{\nu_{\rm receive} - \nu_{\rm send}}{\nu_{\rm send}} \nonumber \\ 
    &\approx& \frac{1}{2\left(1-\bm{\hat{k}} \cdot \bm{\hat{n}}_{ij}(t)\right)}\left[H_{ij}\left(t - \frac{d_{ij}(t)}{c} - \frac{\bm{\hat{k}} \cdot \bm{R}_j(t)}{c}\right) \right. \nonumber \\ 
    && \left.- H_{ij}\left(t - \frac{\bm{\hat{k}} \cdot \bm{R}_i(t)}{c}\right)\right].
\end{eqnarray}
$\bm{R}_i(t)$ denotes the position of S/C$_i$ in the Solar system barycenter (SSB) frame, $d_{ij}(t)$ is the light travel time from S/C$_j$ to S/C$_i$ ($t$ being the time of reception), and $\bm{\hat{n}}_{ij}(t)$ is the unit vector along this arm. 
During realistic detection, these quantities should be derived via orbit determination and inter-S/C ranging~\cite{ranging_sensor_fusion,clock_sync_ltt_calculation}, and they generally deviate from the idealistic equal-arm orbit models as adopted by LISA data challenge (LDC)~\cite{LDC_Radler_manual_v2}. 
To incorporate more realistic orbital dynamics, 
all the models and simulations of this work utilize the numerically simulated Taiji Data Challenge orbit~\cite{TDCII}.
The projection of GW tensor on arm $ij$ reads 
\begin{equation}
    H_{ij}(t) \equiv h_+(t) \zeta_{+, ij}(t) + h_\times(t) \zeta_{\times, ij}(t),
\end{equation}
where 
\begin{eqnarray}
\zeta_{+, ij}(t) &=& \cos (2\psi) \xi_{+, ij}(t) + \sin (2\psi) \xi_{\times, ij}(t), \\ 
\zeta_{\times, ij}(t) &=& -\sin (2\psi) \xi_{+, ij}(t) + \cos (2\psi) \xi_{\times, ij}(t), \\
\xi_{+, ij}(t) &=& \left[\bm{\hat{n}}_{ij}(t) \cdot \bm{\hat{u}}\right]^2 - \left[\bm{\hat{n}}_{ij}(t) \cdot \bm{\hat{v}}\right]^2, \\ 
\xi_{\times, ij}(t) &=& 2 \left[\bm{\hat{n}}_{ij}(t) \cdot \bm{\hat{u}}\right]  \left[\bm{\hat{n}}_{ij}(t) \cdot \bm{\hat{v}}\right].
\end{eqnarray}
The $\zeta$s and $\xi$s are the antenna pattern functions in the source frame and SSB frame, respectively, and $\psi$ stands for the polarization angle. 
For a GW source located at Ecliptic longitude $\lambda$ and Ecliptic latitude $\beta$, the Cartesian coordinate components of unit vectors $\bm{\hat{u}}$, $\bm{\hat{v}}$, and  $\bm{\hat{k}}$ are 
\begin{eqnarray}
    \bm{\hat{u}} &=& \left[\sin \lambda, - \cos \lambda, 0 \right], \\ 
    \bm{\hat{v}} &=& \left[- \sin \beta \cos \lambda, - \sin \beta \sin \lambda, \cos \beta \right], \\ 
    \bm{\hat{k}} &=&  -\left[ \cos \beta \cos \lambda, \cos \beta \sin \lambda, \sin \beta \right].
\end{eqnarray}
To model the waveform of GB, we utilize the Newtonian approximation and expand the phase of GW to the third power of time, consistent with the widely adopted form in the literature (\emph{e.g.} Ref.~\cite{orbit_approximation}):
\begin{eqnarray}
    h_+(t) &=& A (1 + \cos^2 \iota) \cos  \varphi \left(t\right) , \\
    h_\times(t) &=& 2A \cos \iota \sin  \varphi \left(t\right) ,
\end{eqnarray}
where the amplitude  $A$ and phase $\varphi(t)$ are related to the chirp mass $\mathcal{M}_c$, luminosity distance $D$, initial GW phase $\varphi_0$, initial GW frequency $f_0$ and its derivatives as 
\begin{eqnarray}
    A &=& \frac{2(G\mathcal{M}_c)^{5/3}(\pi f_0)^{2/3}}{c^4 D}, \\
    \varphi(t) &=& 2\pi\left(f_0 t + \frac{1}{2}\dot{f}_0 t^2 + \frac{1}{6} \ddot{f}_0 t^3\right) + \varphi_0,
\end{eqnarray}
respectively, with 
\begin{equation}\label{eq:frequency_2nd_derivative}
    \ddot{f}_0 = \frac{11}{3}\frac{\dot{f}_0^2}{f_0}.
\end{equation}

In principle, the noise term $n_{ij}(t)$ of Eq.~(\ref{eq:single_arm_data}) should be regarded as the  synthesis of multiple noises,
such as test-mass acceleration noises, optical path noises, readout noises, and fibre backlink noises, \emph{etc}.
The dominant contributions can be described as a two-component model including the optical metrology system (OMS) noises $N_{ij}$ and test-mass acceleration (ACC) noises $\delta_{ij}$: 
\begin{equation}
    n_{ij}(t) = N_{ij}(t) + \delta_{ij}(t) + \textbf{D}_{ij} \delta_{ji}(t).
\end{equation}
The delay operator $\textbf{D}_{ij}$ is defined as $\textbf{D}_{ij} f (t) \equiv f\left(t - d_{ij}(t) \right)$, when acted on an arbitrary function of time  $f(t)$. 
Readers may refer to Ref.~\cite{noise_transfer_functions} for a more comprehensive and up-to-date model incorporating other  subdominant noise sources. 
According to the baseline design of Taiji, the ``nominal'' PSDs of $N_{ij}$ and $\delta_{ij}$  take the forms of 
\begin{eqnarray}\label{eq:noise_component_PSDs}
    S_{ij, {\rm OMS}}(f) &=& A_{ij, {\rm OMS}}^2 \left(\frac{2\pi f}{c}\right)^2 \left[1 + \left(\frac{2 \ {\rm mHz}}{f}\right)^4\right], \nonumber \\
    S_{ij, {\rm ACC}}(f) &=& A_{ij, {\rm ACC}}^2 \left(\frac{1}{2\pi f c}\right)^2  \left[1 + \left(\frac{0.4 \ {\rm mHz}}{f}\right)^2\right] \nonumber \\ 
    & & \times  \left[1 + \left(\frac{f}{8 \ {\rm mHz}}\right)^4\right], 
\end{eqnarray}
where $A_{ij, {\rm OMS}} \equiv 8 \times 10^{-12} \ \mathrm{m}/\sqrt{\mathrm{Hz}}$ and $A_{ij, {\rm ACC}} \equiv 3 \times 10^{-15} \ \mathrm{m}/\mathrm{s}^2/\sqrt{\mathrm{Hz}}$.
The OMS and ACC noises dominate at the high and low frequencies, respectively. 
Notice that in realistic detection, we cannot expect these PSDs to be identical for all the $ij$s, and they are very likely to drift over time. 

To effectively suppress laser frequency noise, Taiji utilizes the TDI observables rather than  single-arm measurements for  GW detection. 
Despite the variety of TDI configurations~\cite{Tinto_time_2021,Tinto_time_2004,geometric_tdi_combination},
all the TDI observables  can be written in a unified form:
\begin{equation}\label{eq:general_TDI}
    {\rm TDI} \ = \sum_{ij \in \mathcal{I}_2} \textbf{P}_{ij} \eta_{ij}.
\end{equation}
Taking  the Michelson $X_2$ channel as an example, one has
\begin{eqnarray}\label{eq:X2_combination}
    \textbf{P}_{12} &=& 1 - \textbf{D}_{131} - \textbf{D}_{13121} + \textbf{D}_{1213131}, \nonumber \\ 
    \textbf{P}_{23} &=& 0, \nonumber \\ 
    \textbf{P}_{31} &=& -\textbf{D}_{13} + \textbf{D}_{1213} + \textbf{D}_{121313} - \textbf{D}_{13121213}, \nonumber \\ 
    \textbf{P}_{21} &=& \textbf{D}_{12} - \textbf{D}_{1312} - \textbf{D}_{131212} + \textbf{D}_{12131312}, \nonumber \\ 
    \textbf{P}_{32} &=& 0, \nonumber \\ 
    \textbf{P}_{13} &=& -1 + \textbf{D}_{121} + \textbf{D}_{12131} - \textbf{D}_{1312121},  
\end{eqnarray}
where $\textbf{D}_{i_1i_2i_3 ...} f(t) \equiv \textbf{D}_{i_1i_2}\textbf{D}_{i_2i_3}...f(t)$. 
The formulae of $Y_2$ and $Z_2$ channels can be obtained from the permutation rule $1 \rightarrow 2, 2 \rightarrow 3, 3 \rightarrow 1$.
Further,  the ``signal'' channel $A_2, E_2$ and  the ``null'' channel $T_2$ are defined as~\cite{optimal_tdi_channels} 
\begin{eqnarray}\label{eq:AET_from_XYZ}
    A_2 &=& \frac{Z_2 - X_2}{\sqrt{2}}, \\
    E_2 &=& \frac{X_2 - 2Y_2 + Z_2}{\sqrt{6}}, \\
    T_2 &=& \frac{X_2 + Y_2 + Z_2}{\sqrt{3}}. 
\end{eqnarray}

\subsection{\label{subsec:stft_template} The STFT formalism for  Bayesian inference}

Time-frequency domain representations such as STFT~\cite{sft_gw_analysis} and wavelet decompositions~\cite{time_frequency_analysis_cornish}  are generally more suitable than Fourier transform under non-stationary noises~\cite{evolutionary_psd}. 
Meanwhile, data gaps are also inevitable for long-duration scientific operations, either due to scheduled maintenances or unscheduled disruptions. 
Recent studies~\cite{mind_the_gap,glitch_gap} suggested that  gaps in LISA data could occur up to hundreds of times per year, resulting in a duty cycle of $\sim 80\%$ or less~\cite{lisa_mission_duration}. 
To date, the gap profile of Taiji is still not fully characterized, while to also account for its potential impacts, we choose STFT among all time-frequency domain representations, since data is naturally divided into time segments in the presence of gaps.

When adopting STFT, a technical consideration follows, that is, each segment should be windowed to prevent spectral leakage. 
The windowing process also induces  distortion of waveforms, therefore the template used in matched filtering must be identically windowed to ensure unbiased estimation. 
Following practices in ground-based detection and balancing spectral leakage suppression against signal strength preservation, we employ a Tukey window with parameter $\alpha=0.1$. 
Its mathematical expressions in the time and frequency domains are detailed in Appendix~\ref{appendix:tukey_window}.




Consider a generic case where the data is divided 
into $N_t$ segments, with the $m$-th segment centering at $t_m$ and having a length of $T_m$.
Before STFT, each segment is shifted to starting at $t=0$ and multiplied by a window function $w_m(t - T_m/2)$ (since the window is originally defined in $(-T_m/2, T_m/2)$).
Therefore, taking the TDI-$X_2$ channel as an example, the windowed STFT of the $m$-th segment is 
\begin{eqnarray}\label{eq:stft_definition}
\bar{X}_2(t_m, f) &\equiv&  \int_{0}^{T_m} {\rm d}t \ e^{-i2\pi f t} w_m\left(t - \frac{T_m}{2}\right) \nonumber \\ 
 && \times \  X_2\left(t + t_m - \frac{T}{2}\right).
\end{eqnarray}

The derivation of STFT GB template generally leverages the fact that, terms  related to detector orbit and GW frequency are slowly varying, while the GW phase exhibits relatively fast variation~\cite{fastgb}.
Within current models for GB populations (\emph{e.g.} Refs.~\cite{DWD_observation_population,DWD_bps_population}), the evolution of GW frequency is limited to $\dot{f}_0 \lesssim \mathcal{O}(10^{-12})$ Hz/s, thus we can safely approximate $\varphi(t)$ to the linear order during a time span of $\mathcal{O}(10^2)$ s~\footnote{When modeling the  TDI response of GB signal, the amplitude evolves slowly and can be calculated on a  sparse time grid without loss of precision, leaving  the accuracy of phase as our primary focus. To set a criterion for the  accuracy of template,  we require the residual signal-to-noise ratio (SNR) caused by template error to be less than 1. For an order-of-magnitude estimation, we assume a small constant phase error  $\Delta \varphi$ and an original  SNR $\rho_0$. It follows that the SNR of template error approximates $\Delta \varphi \rho_0$. Therefore, for the vast majority of observable GBs, $\Delta \varphi < 10^{-2}$ should suffice to meet the accuracy criterion. The threshold of $\Delta \varphi < 1/\rho$ is also commonly adopted in the literature on EMRI waveforms~\cite{emri_waveform_threshold1,emri_waveform_threshold2}.}, and write $y_{ij}(t)$ as 
\begin{equation}\label{eq:eta_approximation}
    y_{ij}(t) \approx B_{y, ij}(t) e^{i \varphi(t)}  + {\rm c.c.},
\end{equation}
where ``c.c.'' stands for the complex conjugate of the first term, and
\begin{eqnarray}
    B_{y, ij}(t) &\equiv& -\frac{i}{2} \dot{\varphi}(t) d_{ij} {\rm sinc}\left[ \frac{1}{2} \dot{\varphi}(t) d_{ij} \left(1 - \hat{\bm{k}} \cdot \hat{\bm{n}}_{ij}\right) \right] \nonumber \\ 
    & & \times \ e^{-i \dot{\varphi}(t)\left[\frac{d_{ij}}{2} + \frac{\hat{\bm{k}} \cdot \left(\bm{p}_i +\bm{p}_j \right)}{2c} + \frac{\hat{\bm{k}} \cdot \bm{R}_0}{c}\right]} \nonumber \\ 
    & & \times \left(\sum_{\alpha} \zeta_{\alpha, ij} A_\alpha\right), 
\end{eqnarray}
where $\bm{R}_0$ is the center-of-mass of the S/Cs, $\bm{p}_i \equiv \bm{R}_i -\bm{R}_0$, and all the $d_{ij}$, $\hat{\bm{n}}_{ij}$, $\bm{p}_i$ and $\bm{R}_0$ take the values at $t$. 
$A_{\alpha} \ ( \alpha \in \{+, \times\})$ are defined as 
\begin{equation}
    A_+ \equiv \frac{1}{2}A\left(1 + \cos^2 \iota \right), \quad A_\times \equiv -i A \cos \iota. 
\end{equation}
Also due to the slow-variation of GW frequency and detector orbit, each delay operation $\textbf{D}f(t) \equiv f(t - d)$ in Eq.~(\ref{eq:general_TDI}) and Eq.~(\ref{eq:X2_combination}) can be replaced by a factor of $e^{-i \dot{\varphi}(t)d}$, hence simplifying the modeling of TDI combination. 
Considering a typical maximum segment length of $\mathcal{O}(10^5)$ s, 
we further expand the ``fast'' term $\varphi(t)$ around the center of  segment, and neglect the quadratic term (corresponding to an error of less than $\mathcal{O}(10^{-2})$).
By analytically evaluating the integral in Eq.~(\ref{eq:stft_definition}), the resulting GB template reads 
\begin{eqnarray}\label{eq:X2_template}
    \bar{X}_2(t_m, f) \approx e^{-i \pi f T}     \tilde{w}_m\left[f - \frac{\dot{\varphi}(t_m)}{2\pi}\right] A_{X_2}(t_m)
\end{eqnarray}
for $f > 0$, where 
\begin{eqnarray}\label{eq:stft_amplitude}
A_{X_2}(t_m) &=& \left\{ \left(1  - \Delta_{31}^2\right)\left[B_{y, 12}(t_m) + \Delta_{12} B_{y, 21}(t_m)\right] \right. \nonumber \\ 
& & \left. - \left(1  - \Delta_{12}^2\right)\left[B_{y, 13}(t_m) + \Delta_{31} B_{y, 31}(t_m)\right]\right\} \nonumber \\ 
& & \times  \left(1 - \Delta_{12}^2 \Delta_{31}^2\right) e^{i\varphi(t_m)},
\end{eqnarray}
with  $\Delta_{ij} \equiv e^{-i  \dot{\varphi}(t_m) d_{ij}(t_m)}$.
Note that we do not distinguish between $d_{ij}$ and $d_{ji}$ since $f_0|d_{ij} - d_{ji}|$ is typically less than $\mathcal{O}(10^{-4})$~\footnote{The difference between $d_{ij}$ and $d_{ji}$ is crucial for the suppression of laser noise, though.}. 
The templates for $\{Y_2, Z_2\}$ and  $\{A_2, E_2, T_2\}$ channels can be obtained via the permutation rule and Eq.~(\ref{eq:AET_from_XYZ}).

The deduction of time-varying noise spectral model is relatively simpler. 
Assuming that  $N_{ij}$ and $\delta_{ij}$ for different laser links and test-masses do not correlate,  
the noise PSD of a general TDI observable reads  
\begin{equation}\label{eq:time_frequency_noise_model}
    S(t, f) = \sum_{ij} \left[\big|\tilde{\textbf{P}}_{ij}\big|^2 S_{ij, {\rm OMS}} + \big|\tilde{\textbf{P}}_{ij} + \tilde{\textbf{D}}_{ji}\tilde{\textbf{P}}_{ji}\big|^2 S_{ij, {\rm ACC}} \right].
\end{equation}
Generally $\tilde{\textbf{P}}_{ij}$, $S_{ij, {\rm OMS}}$, $S_{ij, {\rm ACC}}$  all depend on $t$ and $f$.
The time dependence of $\tilde{\textbf{P}}_{ij}$ originates from that of $\tilde{\textbf{D}}_{ij}$, \emph{i.e.} 
\begin{equation}
    \tilde{\textbf{D}}_{ij}(t, f) =  e^{-i 2\pi f d_{ij}(t)}.
\end{equation}
To obtain $S(t, f)$ from data, 
the Fourier transform should be calculated over a time segment which is short enough for the noises to be considered  stationary, and long enough to ensure the frequency resolution of signal and cover the sensitive band of Taiji.

\begin{figure}
    \centering
    \includegraphics[width=0.9\linewidth]{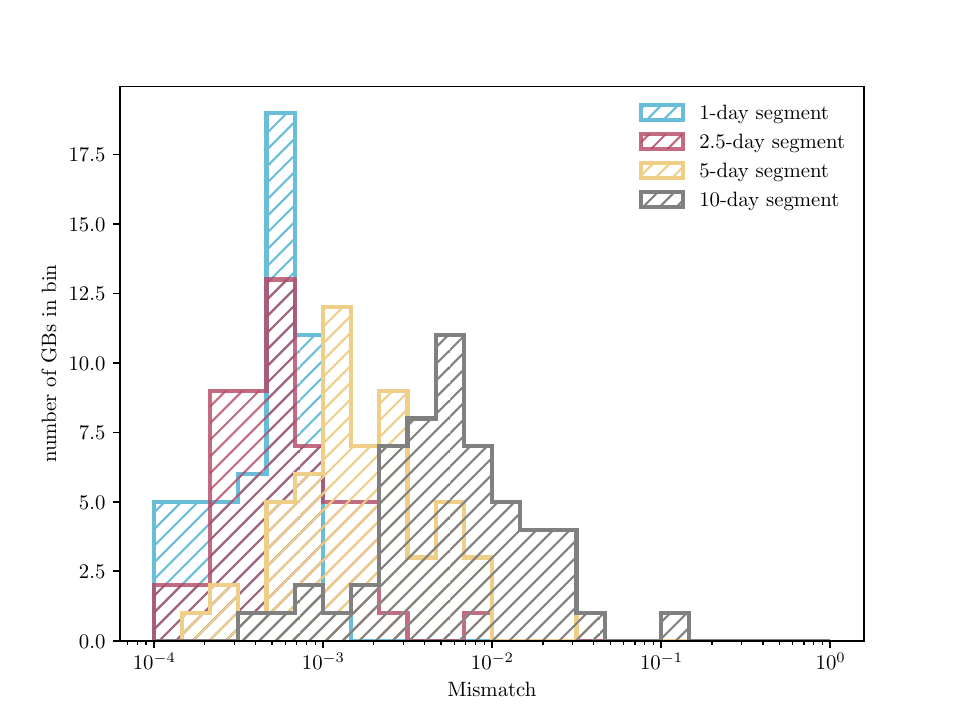}
    \caption{
    Mismatches between STFT templates and rigorous time-domain waveforms.
    The test is performed on 55 VGBs cataloged in Ref.~\cite{Kupfer:2023nqx}. 
    Four distinct lengths of segments are considered: 10 days (grey), 5 days (yellow), 2.5 days (red), and 1 day (blue).
    }
    \label{fig:stft_wf_mismatches}
\end{figure}

Under the conditions of local stationarity  and statistical independence among segments (as verified in Refs.~\cite{mind_the_gap,time_frequency_analysis_cornish}), 
all the time-frequency pixels are uncorrelated, thus 
we define the inner product between data $d$ and template $h$ as 
\begin{equation}
    \langle d | h \rangle \equiv 4 \sum_{m=1}^{Nt}  \sum_{n=1}^{N_{f,m}} \frac{1}{T_m}
    \frac{\Re \left[\bar{d}\left(t_m, f_n\right) \bar{h}^*\left(t_m, f_n\right)\right]}{S\left(t_m, f_n\right)},
\end{equation}
where the frequencies of each STFT depend on $T_m$, \emph{i.e.} $f_n \in \{1/T_m, 2/T_m, ... , N_{f, m}/T_m\}$, with the upper limit $N_{f_m} / T_m$ determined by the Nyquist frequency.
Accordingly, the mismatch $\mathcal{MM}$, a metric quantifying the discrepancy  between waveform $h_1$ and $h_2$, is also extended to STFT representation as  
\begin{equation}
     \mathcal{MM} = 1 - \frac{\langle h_1 |h_2 \rangle}{\sqrt{\langle h_1 | h_1 \rangle \langle h_2| h_2 \rangle}}.
\end{equation}
The validation of our STFT GB template is evaluated using this metric.  
To ensure the accuracy of templates used in the investigation of Section~\ref{sec:gb_estimation}, tests are conducted on the 55 VGBs collected by Ref.~\cite{Kupfer:2023nqx}. 
According to the model in Sec.~\ref{subsec:signal_noise_model}, 
we simulate VGB signals in the time domain using the Taiji Data Challenge toolkit \texttt{Triangle-Simulator}~\footnote{\url{https://github.com/TriangleDataCenter/Triangle-Simulator}}, slice each of them  into uniform segments, perform STFT, and then compare with the templates. 
By varying the lengths of segments, we calculate and compare the resulting mismatches in FIG.~\ref{fig:stft_wf_mismatches}.
Consistent with our  derivation, for segment length of 1 day and 2.5 days ($\sim 10^5$ s), the mismatches are generally at  $\mathcal{O}(10^{-3})$ or $\mathcal{O}(10^{-4})$,  satisfying the requirement for unbiased estimation~\cite{mismatch_threshold1,mismatch_threshold2}. 
Also as expected, increasing the duration would compromise the  accuracy of STFT templates. 
In practice, the selection of $T_m$ should depend on  both the  locations of gaps and  the variations of  noises.
Until further characterizations of Taiji's gap and non-stationary statistics are available,  we currently take  $T_m \equiv 2.5$ days  as a representative example.  


Further, the extended  Whittle likelihood in terms of STFT reads 
\begin{eqnarray}\label{eq:time_frequency_domain_likelihood}
    {\rm ln}\mathcal{L} = - \sum_{m=1}^{Nt}  \sum_{n=1}^{N_{f,m}} && \bigg[ \frac{2}{T_m}
    \frac{  \big| \bar{d}\left(t_m, f_n\right) -\bar{h}\left(t_m, f_n; \bm{\theta}_{\rm signal}\right) \big|^2 }{S\left(t_m, f_n; \bm{\theta}_{\rm noise}\right)} \nonumber \\ 
    && \ \  + \  {\rm ln}S\left(t_m, f_n; \bm{\theta}_{\rm noise}\right) \bigg].
\end{eqnarray}

\section{GB Parameter Estimation}\label{sec:gb_estimation}


As a key component of LISA's prototype global analysis pipelines~\cite{global_fit_Littenberg:2023xpl,global_fit_Strub:2024kbe,global_fit_Katz:2024oqg,global_fit_geemoo},  \texttt{FastGB}~\cite{fastgb}   is currently  the state-of-the-art algorithm for rapid GB signal calculation in the frequency domain. 
In this work, we utilize its GPU implementation \texttt{GBGPU}~\cite{gbgpu1,gbgpu2,gbgpu_zenodo} as a benchmark for the conventional frequency-domain analysis.
Notice that the original \texttt{GBGPU} is based on LISA's equal-arm analytic orbit and  1st-generation TDI.
To stay consistent with our interested scenario,  we  further adapt it to fit Taiji's numerical  orbit and 2nd-generation TDI. 
The adapted version can be found in public GitHub repository \texttt{Triangle-GB}~\footnote{\url{https://github.com/TriangleDataCenter/Triangle-GB}}, 
and its validation against rigorous time-domain simulation is presented in the Appendix of Ref.~\cite{TDCII}.

Realistic space-based GW data analysis is featured by the need to  simultaneously  analyze  and update the models of  signals and noises. 
To emulate this situation without actually performing  a costly global analysis, we designed a ``noise-agnostic'' workflow (distinguished from those that assumes noise spectra are known), and implemented it in both frequency domain and time-frequency domain. 
This enables the comparison of their performances in the presence of non-stationary noises.

\begin{figure}
    \centering
    \includegraphics[width=0.9\linewidth]{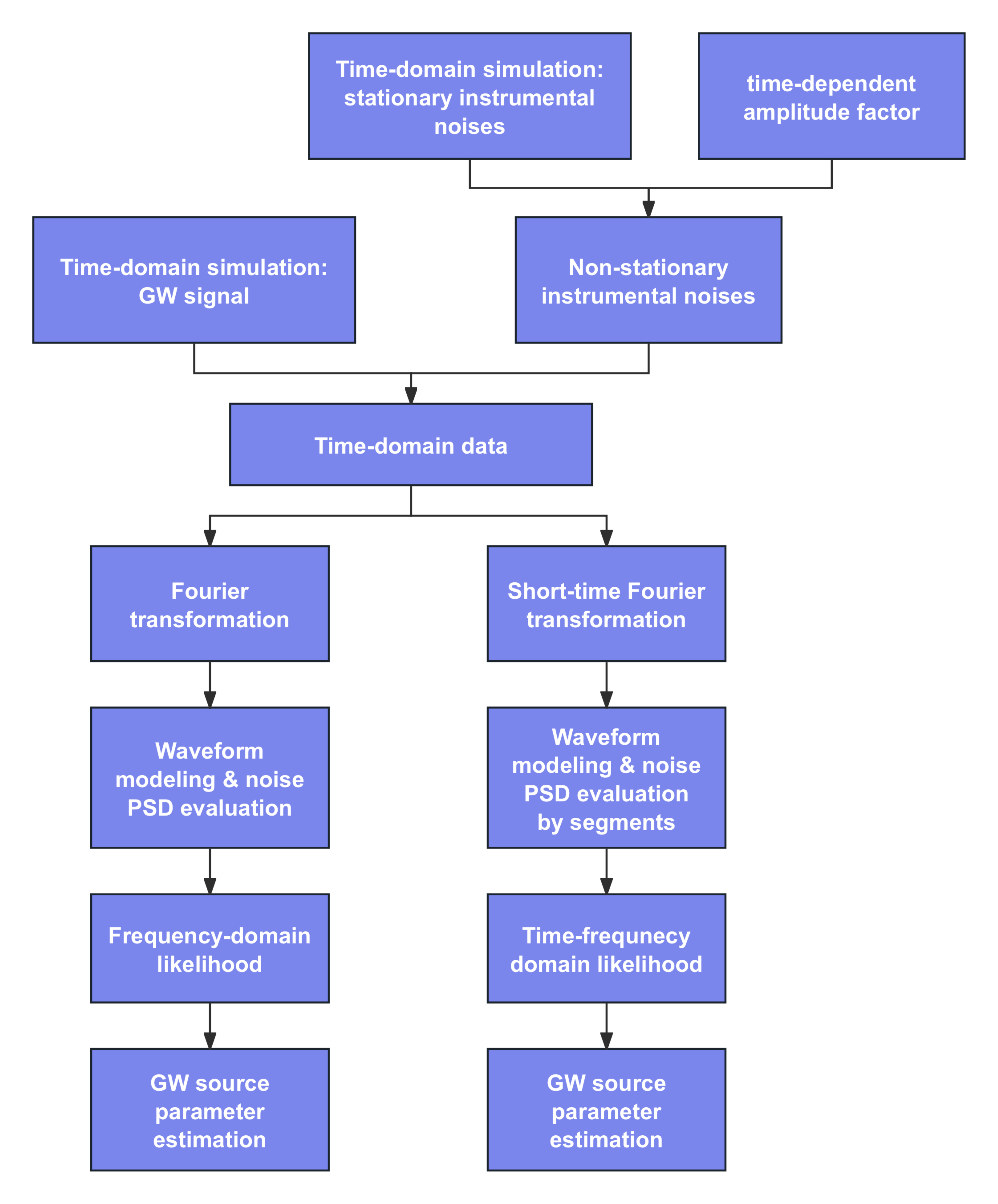}
    \caption{The ``noise agnostic'' workflow.}
    \label{fig:noise_agnostic_workflow}
\end{figure}

\begin{figure}
    \centering
    \includegraphics[width=0.85\linewidth]{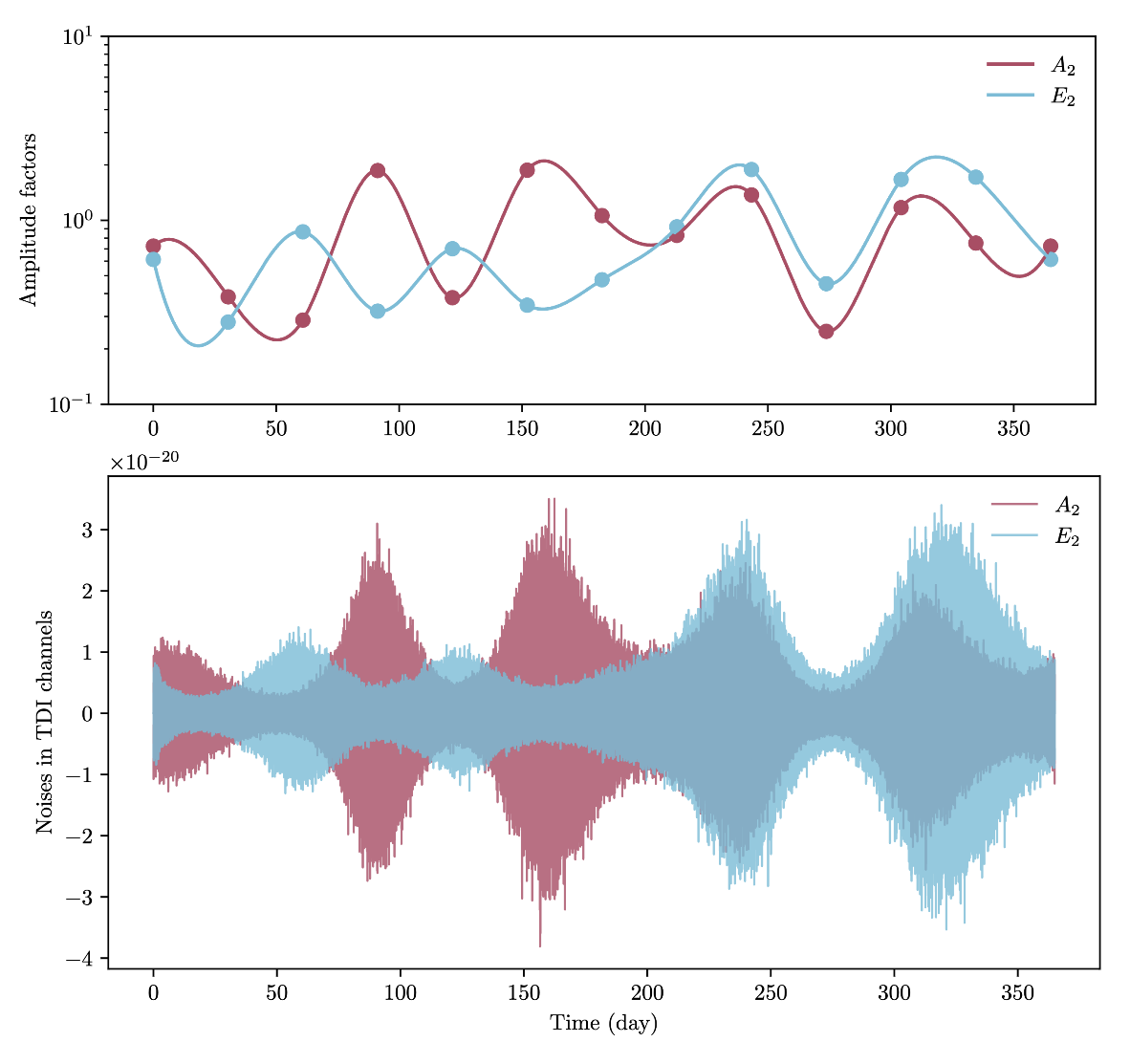}
    \caption{
    Time-varying instrumental noises in the TDI-$A_2, E_2$
    channels (lower panel), obtained by multiplying time-dependent amplitude  factors (upper panel) to the original Gaussian stationary noises simulated with \texttt{Triangle-Simulator}. 
    The amplitude factors, designed to mimic monthly variations, are generated by interpolating from 13 knots using cubic splines. 
    }
    \label{fig:noise_variation}
\end{figure}

As outlined in  FIG.~\ref{fig:noise_agnostic_workflow}, we first simulate signals and noises separately in the form of 2nd-generation TDI variables $\{A_2, E_2, T_2\}$ via \texttt{Triangle-Simulator}. 
For the estimation of GB  parameters, only the  quasi-orthogonal ``signal'' channels $A_2$ and $E_2$ are considered. 
Simulating signals and noises separately  enables the introduction  of time varying   noise amplitude  before  combination. 
Following  the treatment of Ref.~\cite{noise_variation_in_frequency} (but applying to time-dependence  rather than frequency-dependence), we let the noise spectrum vary by $\pm 1$ magnitudes relative to the average level. 
Cubic spline interpolation  with 1 knot per month is adopted to model  a generic slow-varying noise, and the knots for $A_2$, $E_2$ channels are generated independently. 
Subsequently, the noises are further rescaled to ensure their   averaged spectra over a year align with Taiji's nominal PSDs, hence guaranteeing the signal's signal-to-noise ratios (SNRs) are as expected.


\begin{figure*}
    \centering
    \includegraphics[width=0.3\linewidth]{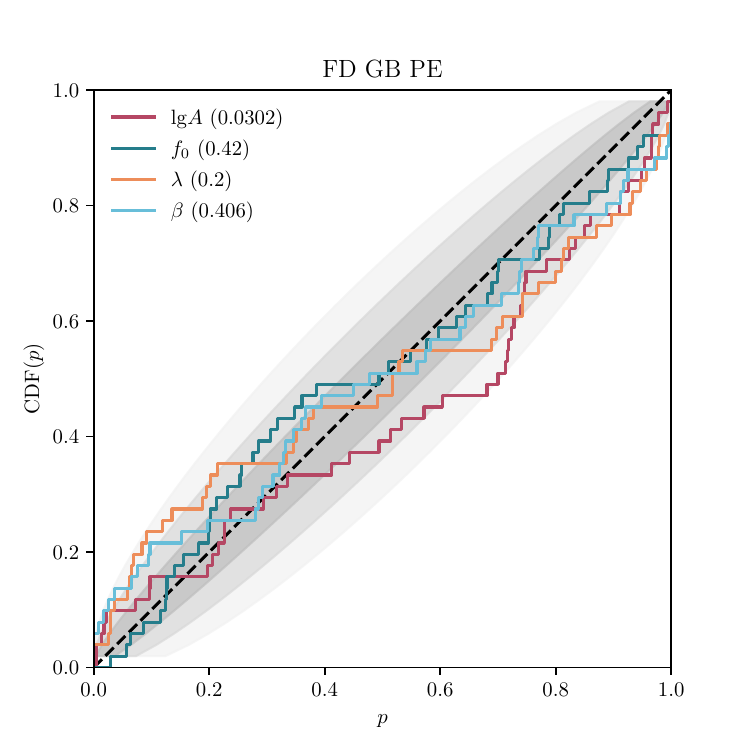}
    \includegraphics[width=0.3\linewidth]{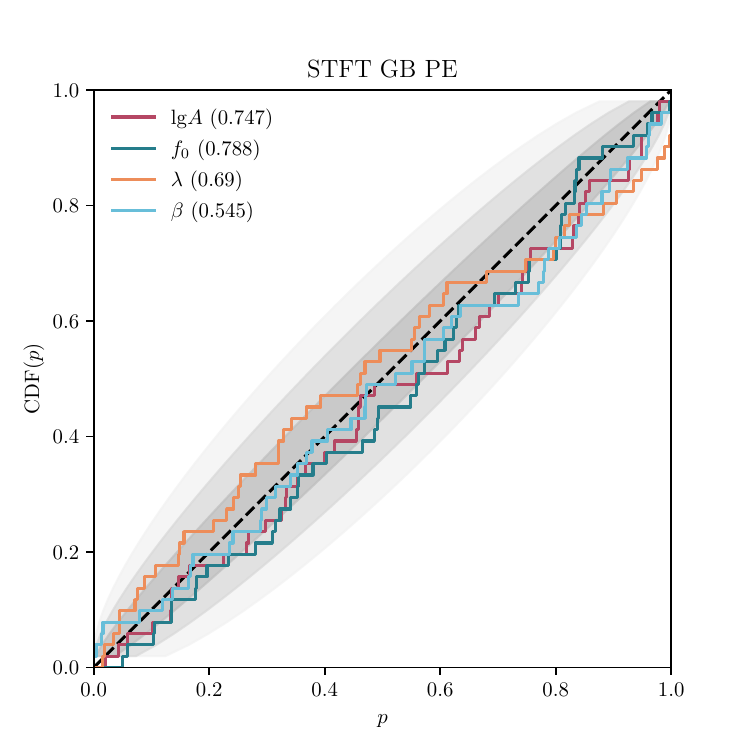} \\
    \includegraphics[width=0.24\linewidth]{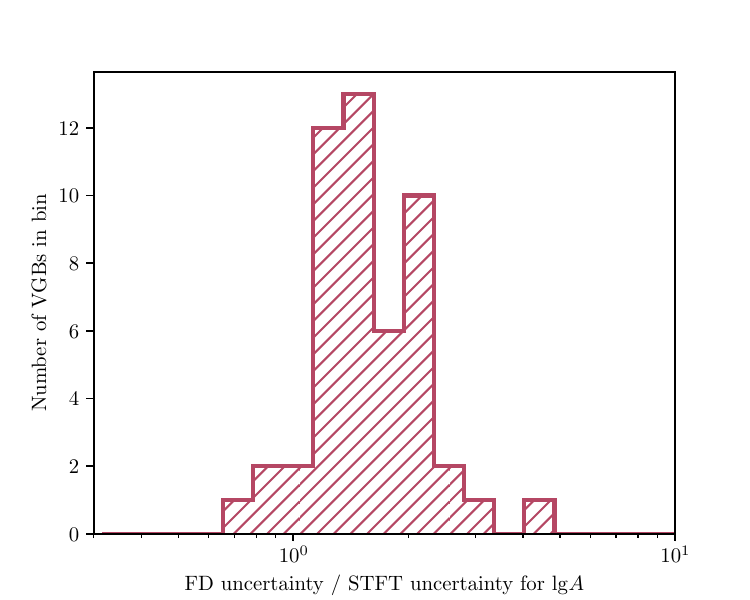}
    \includegraphics[width=0.24\linewidth]{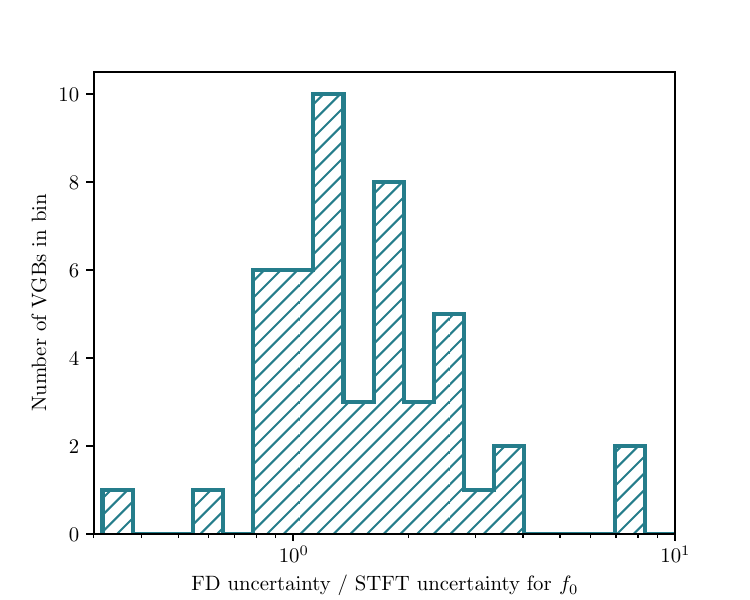}
    \includegraphics[width=0.24\linewidth]{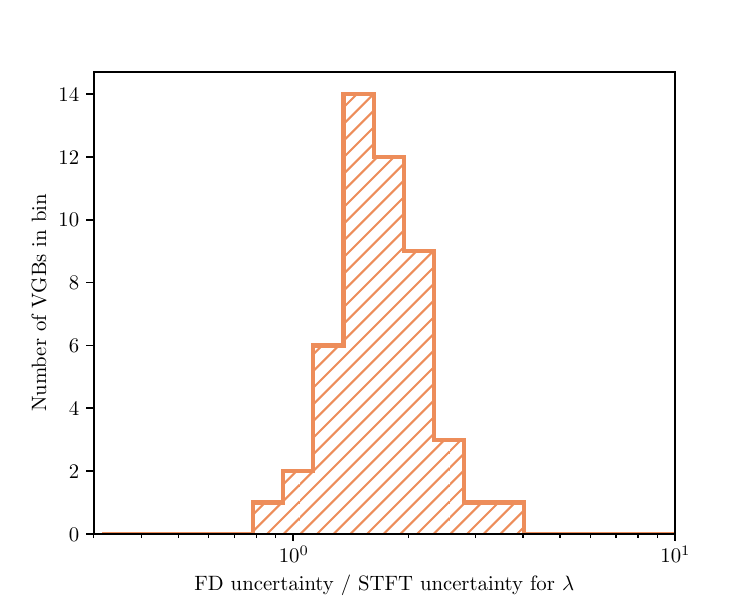}
    \includegraphics[width=0.24\linewidth]{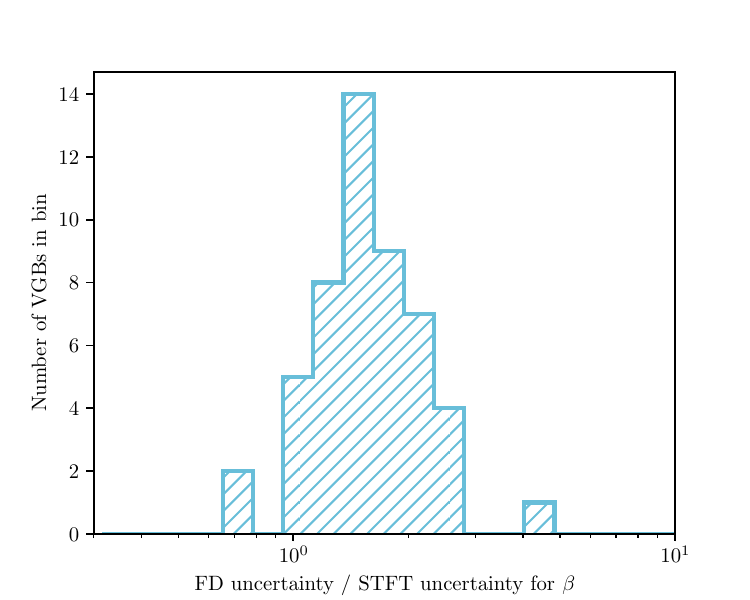} 
    \caption{
    The statistical summaries for four key parameters (amplitude $A$, initial frequency $f_0$, Ecliptic longitude $\lambda$ and latitude $\beta$). 
    The upper panels display the  P-P plots for their  posterior distributions based on frequency-domain and STFT parameter estimations. 
    For each parameter, the
    curve represents the cumulative distribution function of the percentile scores of the true values. 
    The black dashed line indicates ideal unbiased estimate, while  gray regions denote the 1$\sigma$ to 3$\sigma$ confidence bounds. 
    The $p$-values of Kolmogorov-Smirnov (KS) test are provided in the legend. 
    Compared in the lower panels are the uncertainties (\emph{i.e.}, 1$\sigma$ credible interval widths) of parameters. 
    }
    \label{fig:vgb_posterior_statistics}
\end{figure*}

\begin{figure}
    \centering
    \includegraphics[width=0.85\linewidth]{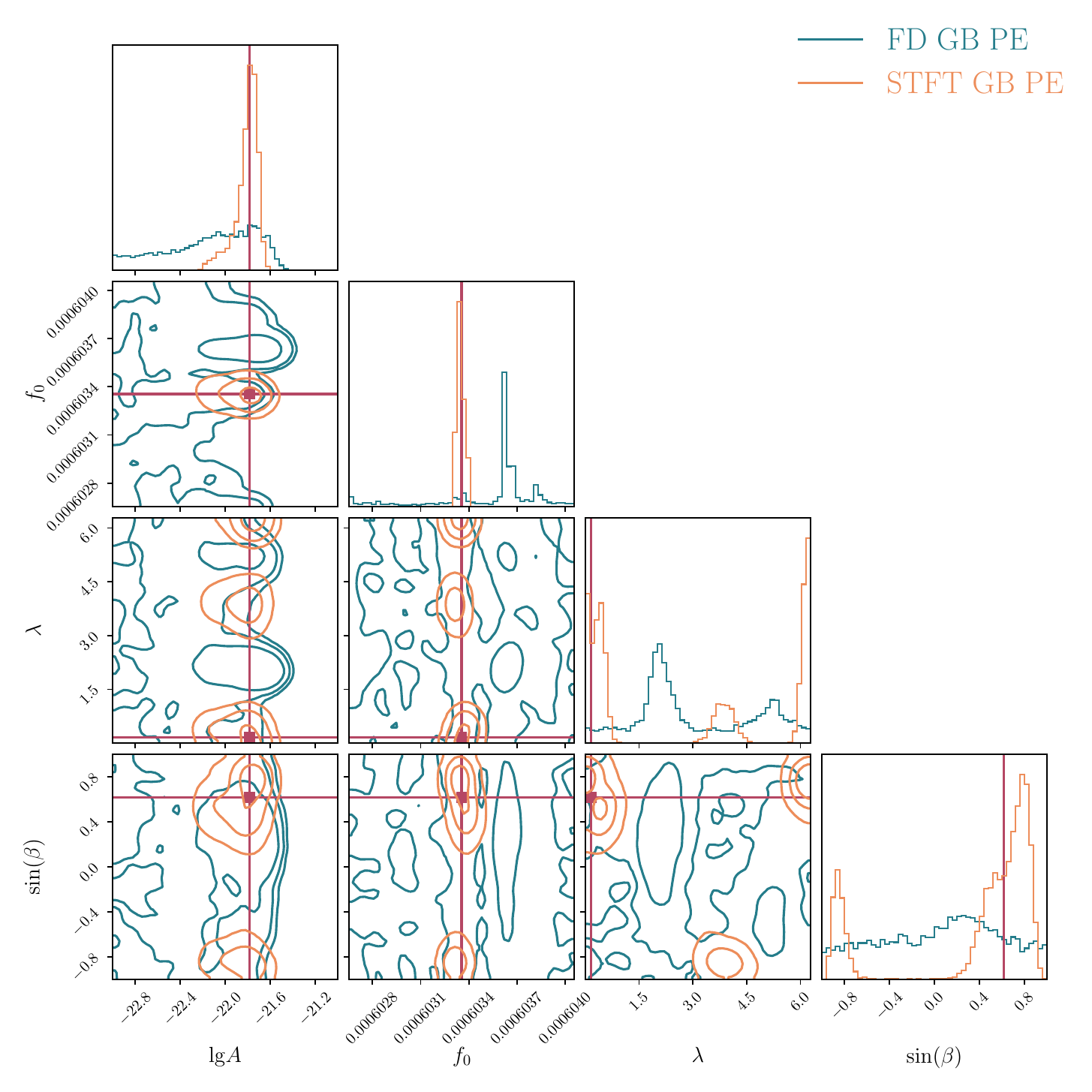}
    \caption{The posteriors of VGB ZTF J2320 as an extreme example for the advantage of STFT analysis (orange)  over frequnecy-domain analysis (green) under noise non-stationarity.}
    \label{fig:ZTFJ2320}
\end{figure}


When working in the frequency domain, a $10 \ \mu{\rm Hz}$ band centered at  the initial frequency $f_0$ of  target signal is used for analysis.
This bandwidth  is a conservative choice to ensure the entire   evolution of signal is accounted for. 
Evaluating noise PSD in the frequency domain is straightforward: we  calculate the periodogram within this  band and  take the median values to avoid  the disruption caused by  signal~\cite{find_chirp,global_fit_Strub:2024kbe} (neglecting the spectral shape due to the narrow bandwidth).
On the other hand, in time-frequency domain, to capture  the time dependence of PSD (\emph{i.e.}  to get $S(t, f_0)$), we select $50 \ \mu{\rm Hz}$ around $f_0$ (for the relatively coarse frequency resolution, also conservatively), calculate the median over this frequency band  for each time segment,  and further  smooth the resulting $S(t, f_0)$ time series with the Savitzky-Golay  filter~\cite{savgol_filter}. 

Given the noise spectra, one can neglect all the $\bm{\theta}_{\rm noise}$ in Eq.~(\ref{eq:time_frequency_domain_likelihood}), thus only the first line remains. 
For each TDI channel, 
\begin{eqnarray}\label{eq:time_frequency_domain_signal_likelihood}
    {\rm ln}\mathcal{L} = - \sum_{mn} &&  \frac{2}{T_m}
    \frac{  \big| \bar{d}\left(t_m, f_n\right) -\bar{h}\left(t_m, f_n; \bm{\theta}_{\rm signal}\right) \big|^2 }{S\left(t_m, f_n\right)},
\end{eqnarray}
and the total log likelihood is obtained by summing over $A_2$ and $E_2$. 
Neglecting the dependence on  $t$, above expression reduces to the conventional frequency-domain formula used for comparison, where the term ``conventional'' suggests that correlations among different frequencies are not accounted for. 

Tests are conducted on 55 VGBs following the ``noise agnostic'' workflow. 
For each VGB, both the noise realization and amplitude knots are regenerated. 
Notice that these VGBs are defined according to their detectability  under 48-month observations~\cite{Kupfer:2023nqx}, 
whereas the time span of our simulation is only 1 year. 
To ensure sufficient  SNR for each source, the amplitudes of the first 25 sources (ordered per Ref.~\cite{VGB_gitlab}) are amplified by a factor of 2, and the remaining 30 by a factor of 4.
Markov Chain Monte Carlo (MCMC) sampling of the likelihood is conducted via  the \texttt{Eryn} sampler~\cite{eryn} with affine invariant move, and the hyperparameters are set as \texttt{n\_walker}=200, \texttt{n\_temperature}=4.

For instance, Appendix~\ref{appendix:posterior} displays the results for two  VGBs: one is HMCnc at high frequency ($f_0=$ 6.22 mHz), and the other is ZTF J1905 at mid-low frequency ($f_0=$ 1.94 mHz).
The posteriors drawn from frequency-domain and STFT likelihoods are visualized in green and orange corner plots, respectively. These results preliminarily demonstrate the advantage of time-frequency analysis under non-stationary noise, as evidenced by  reduced uncertainties and biases. 
Extending to more comprehensive comparison across all sources, FIG.~\ref{fig:vgb_posterior_statistics} presents the statistical summaries of four key parameters (amplitude $A$, initial frequency $f_0$, Ecliptic longitude $\lambda$ and latitude $\beta$)~\footnote{For parameters with multimodal distributions (\emph{e.g.}, the sky locations of some sources), we employ \texttt{SCIPY}'s \texttt{find\_peaks} method to select only the peak with maximum likelihood values}. 
To compare the statistical unbiasedness of the two methods, the  P-P plots for their posterior distributions are shown in the upper panels of FIG.~\ref{fig:vgb_posterior_statistics}. 
For the posteriors of each parameter, the
curve represents the cumulative distribution function of the percentile scores of the true values. 
The black dashed line indicates ideal unbiased estimate, while  gray regions denote the 1$\sigma$ to 3$\sigma$ confidence bounds. 
The $p$-values of Kolmogorov-Smirnov (KS) test are provided in the legend. 
Evidently, both methods yield statistically unbiased results within the 3$\sigma$ level, while the performance of STFT approach  marginally  superior across all the parameters of interest.
Beyond unbiasedness tests, we further compare the parameter constraint capabilities of both methods by computing the ratio of their uncertainties (\emph{i.e.}, 1$\sigma$ credible interval widths). 
Results are shown in the lower panels of FIG.~\ref{fig:vgb_posterior_statistics}. 
These comparisons confirm our previous conclusion drawn from the two VGBs: under non-stationary noise, the STFT approach generally  yields more tightly constrained source parameters. 
Additionally, an extreme case (ZTF J2320, FIG.~\ref{fig:ZTFJ2320}) reveals that  frequency-domain analysis may fail to detect relatively low-SNR signals under non-stationary noise (the signal is almost ``lost''),  while  it is successfully  identified by STFT. 
This underscores that adopting appropriate data representations helps to ensures  faithful characterization of detectability.



\section{\label{sec:noise_characterization} instrumental noise characterization}

Accurate and comprehensive knowledge on instrumental noises is crucial for  assessing the detector's operation status, as well as  detecting resolvable and stochastic signals. 
Among all the configurations for TDI, the so called ``null'' channels are particularly significant  for this purpose, because  they   are insensitive to GWs and mainly carry information about instrumental noises~\cite{sagnac_null_channels,null_channels_PhysRevD.105.023009,null_channels_PhysRevD.105.062006,null_channels_PhysRevD.107.082004}. 
Within the widely adopted  Michelson-type TDI channels, the $T$ channel represents  such a null configuration~\cite{null_channels_PhysRevD.105.023009}. 
However,  its effectiveness as a null channel is hindered by two factors. 
Firstly, for realistic unequal-armlength orbits, the $T$ channel have comparable sensitivity with the signal channels at frequencies below 1 mHz and above 0.1 Hz~\cite{unknown_noise_frequency_shape_Baghi_2023,unequal_arm_unequal_amplitude_PhysRevD.107.123531,unequal_arm_lisa_taiji}. 
Secondly, even ideally assuming  the spectra  $S_{ij, {\rm OMS}}$ and $S_{ij, {\rm ACC}}$   remain invariant through the whole mission lifetime, the resulting $S_{T_2}$ for $T$ channel still exhibits significant temporal instability   since it is sensitive to   armlength variations,  making noise estimation in the frequency domain infeasible~\cite{hybrid_relay_noise_characterization}. 
Regarding the first limitation, although  the $T$ channel is not perfectly  immune to signals, its reduced  sensitivity still   permits early-stage decoupling from the global analysis, namely  it can be used as  noise monitor after the few brightest signals have been  subtracted. 
While the second one is the main  focus of this subsection. 
We demonstrate that, given the well measured armlengths (to   nanosecond precision~\cite{ranging_sensor_fusion,clock_sync_ltt_calculation}) and clearly known transfer functions form  $S_{ij, {\rm OMS}}$ and $S_{ij, {\rm ACC}}$  to $S_{T_2}$, 
the time-variation of $S$ induced by   armlength can be actually well modeled, hence enabling  noise  estimation  in the time-frequency domain. 
Crucially, incorporating these modeled temporal information further relieves degeneracies among the noise parameters.

\begin{figure}
    \centering
    \includegraphics[width=0.9\linewidth]{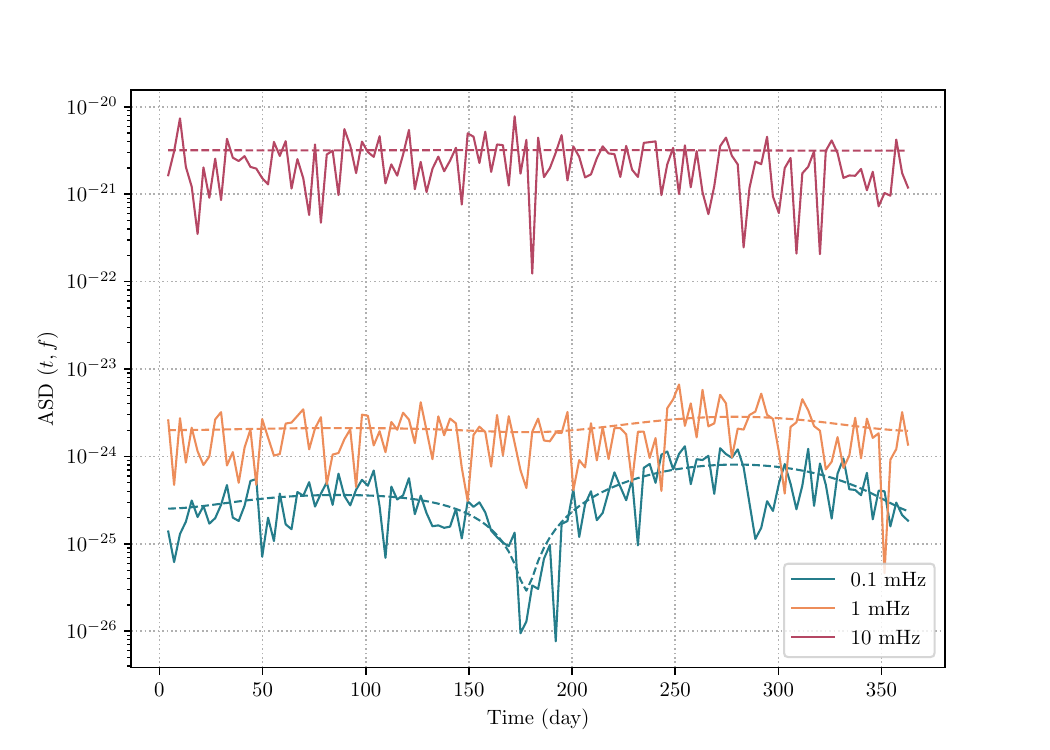}
    \caption{Comparison between the $T$ channel's noise model (dashed curves) and simulated data (solid curves) at 3 representative frequencies $\{0.1 \ {\rm mHz}, 1 \ {\rm mHz}, 10 \ {\rm mHz}\}$.}
    \label{fig:stft_noise_data_vs_model}
\end{figure}

\begin{figure*}
    \centering
    \includegraphics[width=0.49\linewidth]{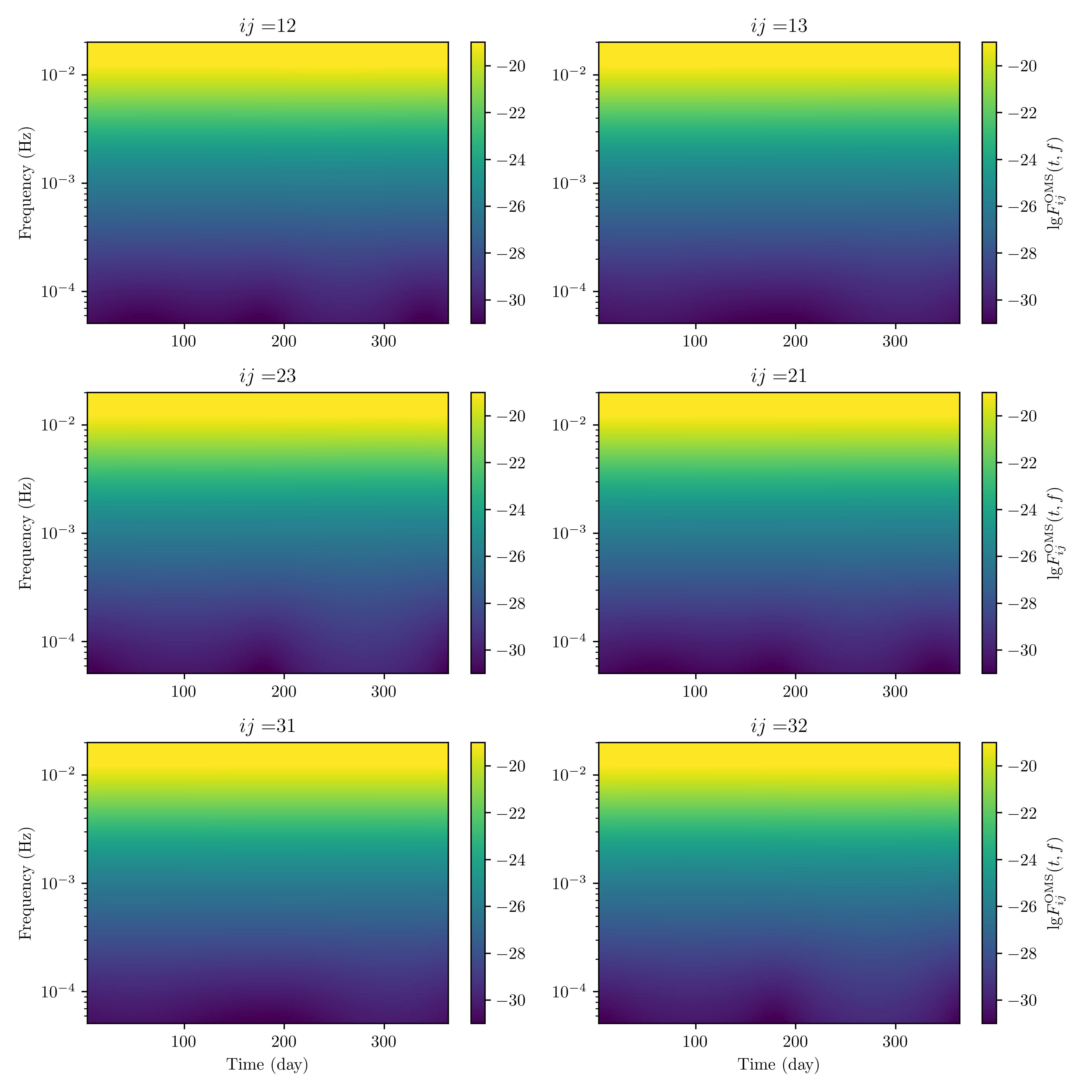}
    \includegraphics[width=0.49\linewidth]{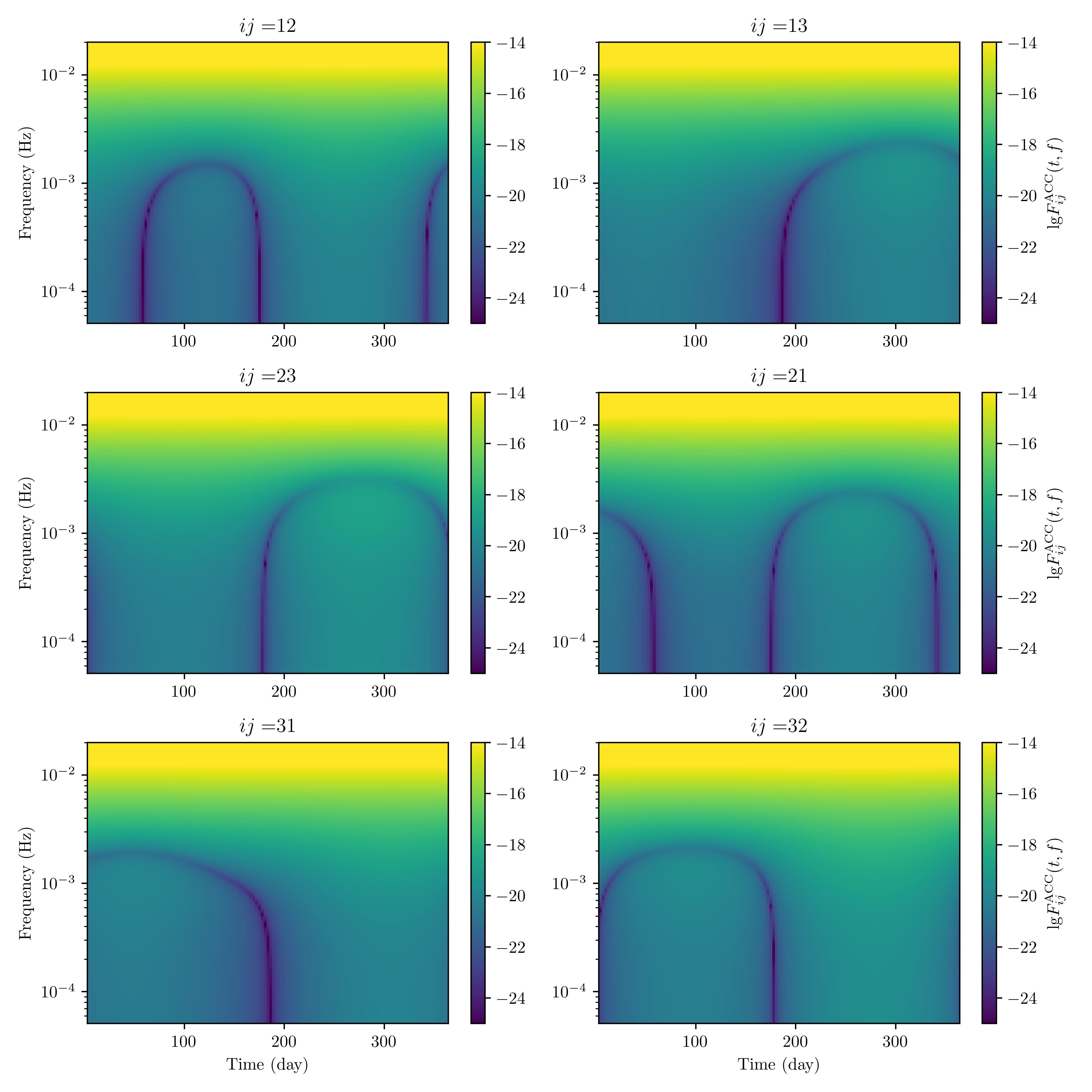}
    \caption{The time-frequency dependence of contributors $F_{ij, {\rm OMS}}$ (left 6 subplots) and $F_{ij, {\rm ACC}}$ (right 6 subplots). }
    \label{fig:stft_noise_contributors}
\end{figure*}

To focus on  the effects of  armlength variation, we first make some conventional  simplifications as  in previous  studies (\emph{e.g.} Refs.~\cite{null_channels_PhysRevD.107.082004,unequal_arm_unequal_amplitude_PhysRevD.107.123531,noise_SGWB_separation_PhysRevD.103.103529,global_fit_Katz:2024oqg,gang_wang_noise_char_1_PhysRevD.106.044054,hybrid_relay_noise_characterization}): 
The spectral shapes of $S_{\rm OMS}$ and $S_{\rm ACC}$ are  known and identical across all $ij \in \{12, 23, 31, 21, 32, 13\}$ (following Eq.~(\ref{eq:noise_component_PSDs})). 
Only their amplitudes $A_{ij, {\rm OMS}}$ and $A_{ij, {\rm ACC, ij}}$ are to be estimated from data, resulting in  $6 \times 2 = 12$ parameters $\bm{\theta}_{\rm noise} = \{A_{ij, {\rm OMS}}, A_{ij, {\rm ACC}}\}$. 
Furthermore,  $S_{ij, {\rm ACC}}$ and $S_{ij, {\rm OMS}}$ are assumed constant throughout the mission lifetime, thereby attributing all time-dependence of $S_{T_2}$ to armlength variation. 
Notably, the noises in realistic  detection scenarios are surely more complicated, thus more flexible spectral modeling such as splines~\cite{unknown_noise_frequency_shape_Baghi_2023,spline_rjmcmc_noise_tianqin} or Gaussian processes~\cite{gaussian_process_lisa_noise_sgwb} remains topic for future research.

\begin{figure*}
    \centering
    \includegraphics[width=0.43\textwidth]{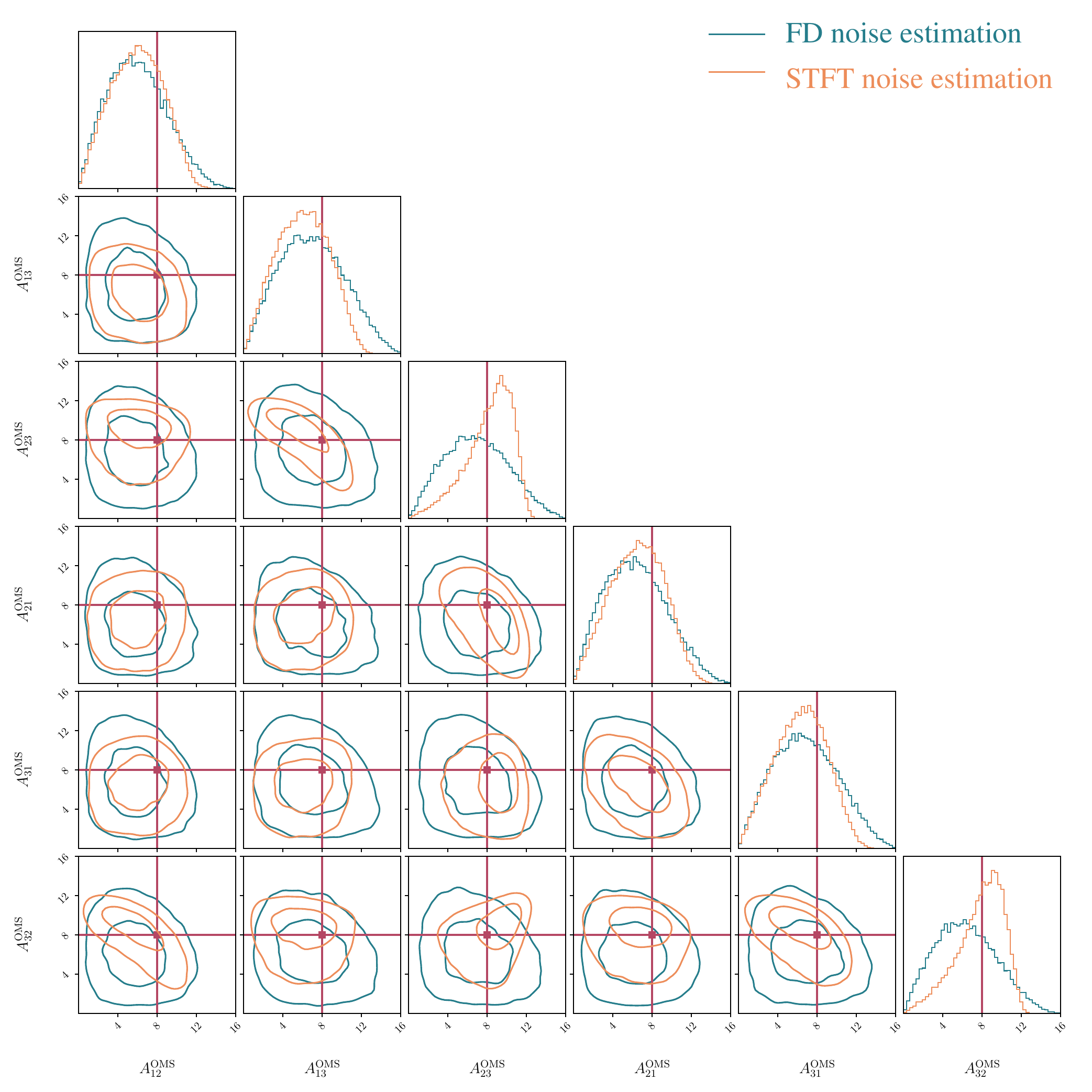}
    \includegraphics[width=0.4\textwidth]{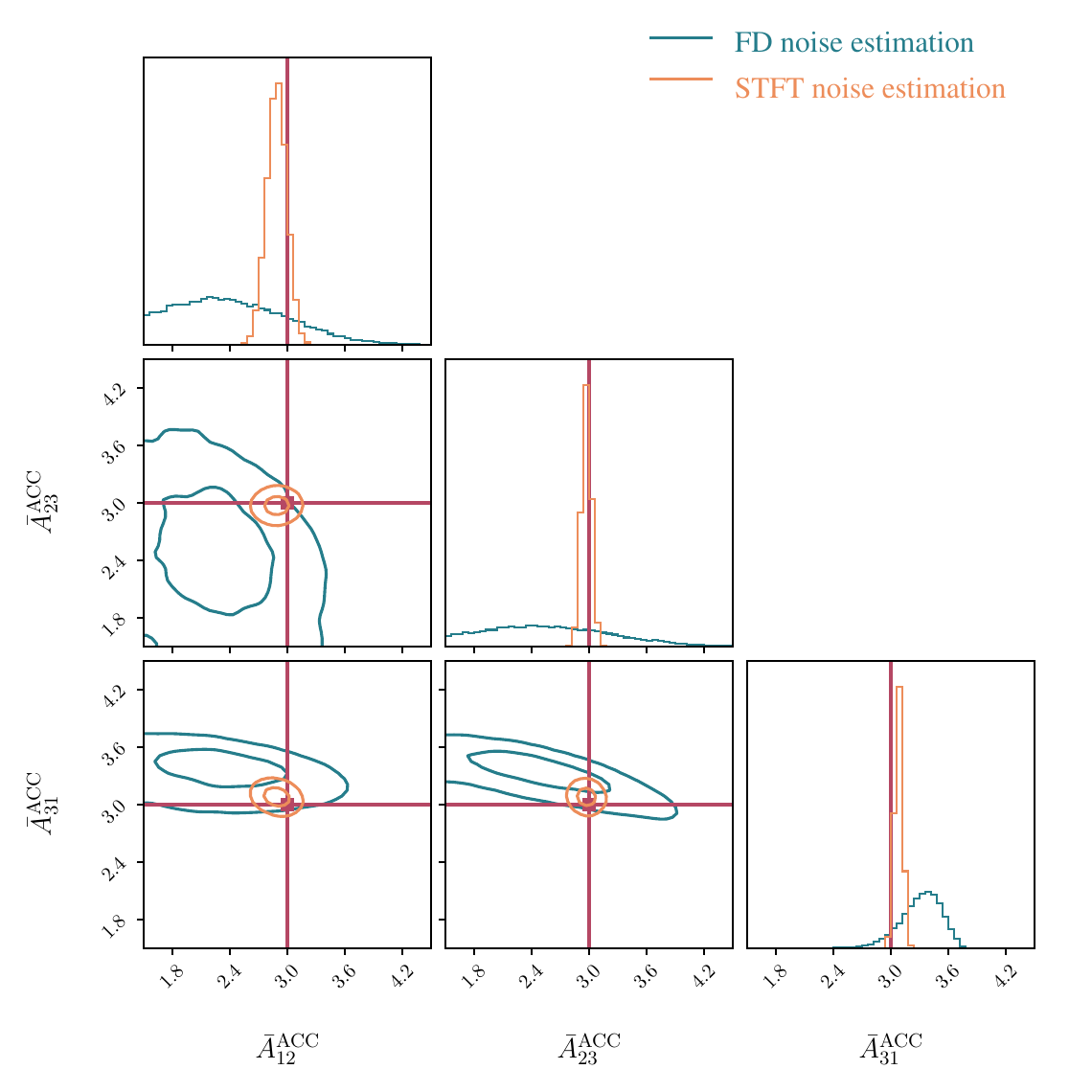} 
    \caption{The posterior distributions of OMS (left) and ACC (right) noise amplitudes. 
    The ACC noise amplitudes are grouped pairwise as  $\bar{A}_{ij, {\rm ACC}}\equiv \sqrt{(A_{ij, {\rm ACC}}^2 + A_{ji, {\rm ACC}}^2) / 2}$.}
    \label{fig:noise_amplitude_posteriors}
\end{figure*}


The arms of Taiji change with a timescale of months (see FIG.~10 of Ref.~\cite{TDCII}). 
Therefore, for the 2.5-day segmentation selected in Section~\ref{subsec:stft_template}, 
we can safely employ the time-frequency domain noise models of Eq.~(\ref{eq:time_frequency_noise_model}) and likelihood function of Eq.~(\ref{eq:time_frequency_domain_likelihood}) under  the local stationarity condition. 
For later convenience, we rewrite the PSD of  $T_2$ channel as
\begin{equation}\label{eq:T_channel_noise_model}
    S_{T_2}(t, f; \bm{\theta}_{\rm noise})  = \sum_{ij, \alpha} A_{ij, \alpha}^2 F_{ij, \alpha}(t, f), 
\end{equation}
where $\alpha \in \{{\rm OMS}, {\rm ACC}\}$. 
The time-frequency dependent contributors $F_{ij, \alpha}$ are calculated from the delay operators.

In FIG.~\ref{fig:stft_noise_data_vs_model}, we select three representative frequencies $\{0.1 \ {\rm mHz}, 1 \ {\rm mHz}, 10 \ {\rm mHz}\}$ to compare the noise model of Eq.~(\ref{eq:T_channel_noise_model}) with the time-varying PSD  calculated from the time-domain simulations of  \texttt{Triangle-Simulator} (we actually show the amplitude spectral densities (ASDs), defined as the square root of PSDs). 
Besides verifying  their consistency, the figure also shows  that the time-dependence of $S_{T_2}(t, f)$  is particularly pronounced at low frequencies.
Further, the time-frequency dependencies of $F_{ij, {\rm OMS}}(t, f)$ and $F_{ij, {\rm ACC}}(t, f)$ are plotted in the left 6 and right 6 subplots of FIG.~\ref{fig:stft_noise_contributors}, respectively. 
Most notably,  the  temporal pattern of $F_{ij, {\rm ACC}}$ are distinguished across different $ij$. 
Given the functional form of Eq.~(\ref{eq:T_channel_noise_model}), this feature  will help to  mitigate  the  degeneracies among parameters, especially $A_{ij, {\rm ACC}}$.

Beyond providing a more rigorous theoretical foundation, we also perform MCMC search for noise parameters using the following noise-only likelihood: 
\begin{eqnarray}\label{eq:time_frequency_domain_noise_likelihood}
    {\rm ln}\mathcal{L} &=& - \sum_{mn} \bigg[ \frac{2}{T_m}
    \frac{  \big| \bar{d}\left(t_m, f_n\right)  \big|^2 }{S_{T_2}\left(t_m, f_n; \bm{\theta}_{\rm noise}\right)}  \nonumber \\ 
      && \quad \quad \quad  \ +  \  {\rm ln}S_{T_2}\left(t_m, f_n; \bm{\theta}_{\rm noise}\right) \bigg]. 
\end{eqnarray}
Notice that since $F_{ij,\alpha}$ are independent of the parameters, they can be precomputed before running MCMC (see the \texttt{TDISensitivity} class of \texttt{Triangle-Simulator} for one implementation). 
As a result, each likelihood reduces to simple operations on  known matrices. 
Besides, to avoid the oscillating behaviors near the ``null'' frequencies~\cite{Wang:2024alm} ($\sim 0.025N$ Hz for Taiji, $N$ = 1, 2, 3, ...), only the frequency range $5 \times 10^{-5}$ - $2 \times 10^{-2}$ Hz is used. 
By neglecting the time dependence, above expression reduces to the frequency-domain version. 
For frequency-domain estimation, we set the model for $S_{T_2}(f)$  as $S_{T_2}(t,f)$ averaged over the whole year. 
The configurations for MCMC sampler and hyperparameters stay  the same as Section~\ref{sec:gb_estimation}. 
In line with the conventions of Refs.~\cite{hybrid_relay_noise_characterization,chinese_knot_tdi}, we present the posterior distributions of OMS and ACC noise amplitudes  separately, and the ACC noise amplitudes are grouped pairwise as  $\bar{A}_{ij, {\rm ACC}}\equiv \sqrt{(A_{ij, {\rm ACC}}^2 + A_{ji, {\rm ACC}}^2) / 2}$
due to the strong intra-pair correlations (\emph{i.e.} $ij$ vs $ji$. See the explanation in Ref.\cite{hybrid_relay_noise_characterization}, and the sub-mHz section of $F_{ij, {\rm ACC}}$ in FIG.~\ref{fig:stft_noise_contributors}). 
FIG.~\ref{fig:noise_amplitude_posteriors} compares the results of frequency-domain (green) and STFT (orange) analysis, with left and right panels showing the posterior distributions of $\bar{A}_{ij, {\rm ACC}}$ and $A_{ij, {\rm OMS}}$, respectively. 
Consistent with our theoretical predictions, by moving to time-frequency domain, the incorporation of modeled temporal features significantly mitigates the degeneracies among the ACC noise parameters, resulting in substantially improved constraints. 
For OMS noise parameters, a modest reduction in the uncertainty  is also observed.






\section{\label{sec:conclusion}conclusion and outlook}
This work develops a STFT-based Bayesian framework to address non-stationary noise in the long-duration GW data analysis of Taiji. 
Focusing on two significant tasks—GB parameter estimation and instrumental noise characterization—we derive GB waveform templates and time-varying noise spectral models, and further  extend Bayesian inference statistics to the STFT domain.  
Compared with conventional frequency-domain methods, our framework demonstrates significantly enhanced accuracy and reduced bias of parameters, confirming the potential of time-frequency representations for the future analysis of realistic Taiji data.

Despite these advantages, certain limitations of the current work warrant further investigation:  

Firstly, while the test on 55 individual VGBs confirms the advantages  of our method, signal overlap remains the core challenge for Taiji data analysis, requiring integrating the time-frequency framework into the whole global analysis pipeline. 
The relatively sparse frequency resolution of STFT  may exacerbate the overlap of GB signals, hence increasing algorithmic complexity and computational demands. 
Nevertheless, results in Section~\ref{sec:gb_estimation} highlight STFT’s critical role in improving the constraints on  source parameters. 
Moreover, its more rigorous theoretical foundation  under noise non-stationarity also ensures the interpretability and credibility of further scientific implications. 
Computationally, our \texttt{CUPY}-accelerated implementation  achieves $\sim 10^4$ likelihood evaluations per second on a NVIDIA V100 GPU, showing potential scalability for more intensive calls by the global fit pipeline. 
Of course, before proceeding to global analysis, further tests on more extended source parameters and  optimization in terms of efficiency and robustness are still required.

Secondly, our noise characterization with $T$ channel assumes non-stationarity arises solely from armlength variations. 
Future work will incorporate the intrinsic drifts of instrumental noise components (OMS and ACC noises, \emph{etc}). 
Besides, realistic detection would involve more complex or unknown noise spectral profiles, necessitating more flexible representations such as splines~\cite{unknown_noise_frequency_shape_Baghi_2023,Li:2025eog,Littenberg:2014oda} and Gaussian process~\cite{gaussian_process_lisa_noise_sgwb}.

\begin{acknowledgments}
This work is supported by National Key Research and Development Program of China
(Grant No. SQ2024YFC220046, No. 2021YFC2201903,
No. 2021YFC2201901, No. 2020YFC2200100).
We gratefully acknowledge He Wang  at University of Chinese Academy of Sciences and Shichao Wu at Max-Planck-Institut f\"{u}r Gravitationsphysik for invaluable discussions and  suggestions.
\end{acknowledgments}

\nocite{*}

\bibliographystyle{apsrev4-2}
\bibliography{apssamp}

\appendix

\section{Tukey window in the time and frequency domain}\label{appendix:tukey_window}
For a Tukey window specified by parameter $\alpha \in [0, 1]$, its time-domain expression reads 
\begin{equation}
    w_T(t) = \begin{cases}
    \frac{1}{2} - \frac{1}{2}\cos\left[\frac{2\pi \left(t + T/2\right)}{\alpha T}\right], &  -\frac{T}{2} \leq t < \frac{\alpha - 1}{2}T \\
    1, & \frac{\alpha - 1}{2}T \leq t < \frac{1 - \alpha}{2}T \\
    \frac{1}{2} - \frac{1}{2}\cos\left[\frac{2\pi \left(T/2 - t\right)}{\alpha T}\right], & \frac{1 - \alpha}{2}T \leq t \leq \frac{T}{2}
\end{cases}
\end{equation}
Applying Fourier transform yields its frequency-domain representation:
\begin{eqnarray}
    \tilde{w}_T(f) &=& \frac{i A^{-1} \left(2B + A - 1\right) - iC^2 \left[1 + A\left(2B -1 \right)\right]}{4 \pi  C \left(B - 1\right) f} \nonumber \\ 
    && + \   \frac{i\left(A^2 C^2 - 1\right)}{2\pi  AC f},
\end{eqnarray}
where we have defined 
\begin{equation}
    A \equiv e^{i\pi  \alpha T f}, \quad  B \equiv  \alpha^2 T^2 f^2, \quad 
    C \equiv e^{-i\pi T f}.
\end{equation}

\begin{widetext}
\section{The full posteriors of VGB parameters based on frequency-domain and STFT analyses}
\label{appendix:posterior}

\begin{figure}[h]
    \centering
    \includegraphics[width=0.8\linewidth]{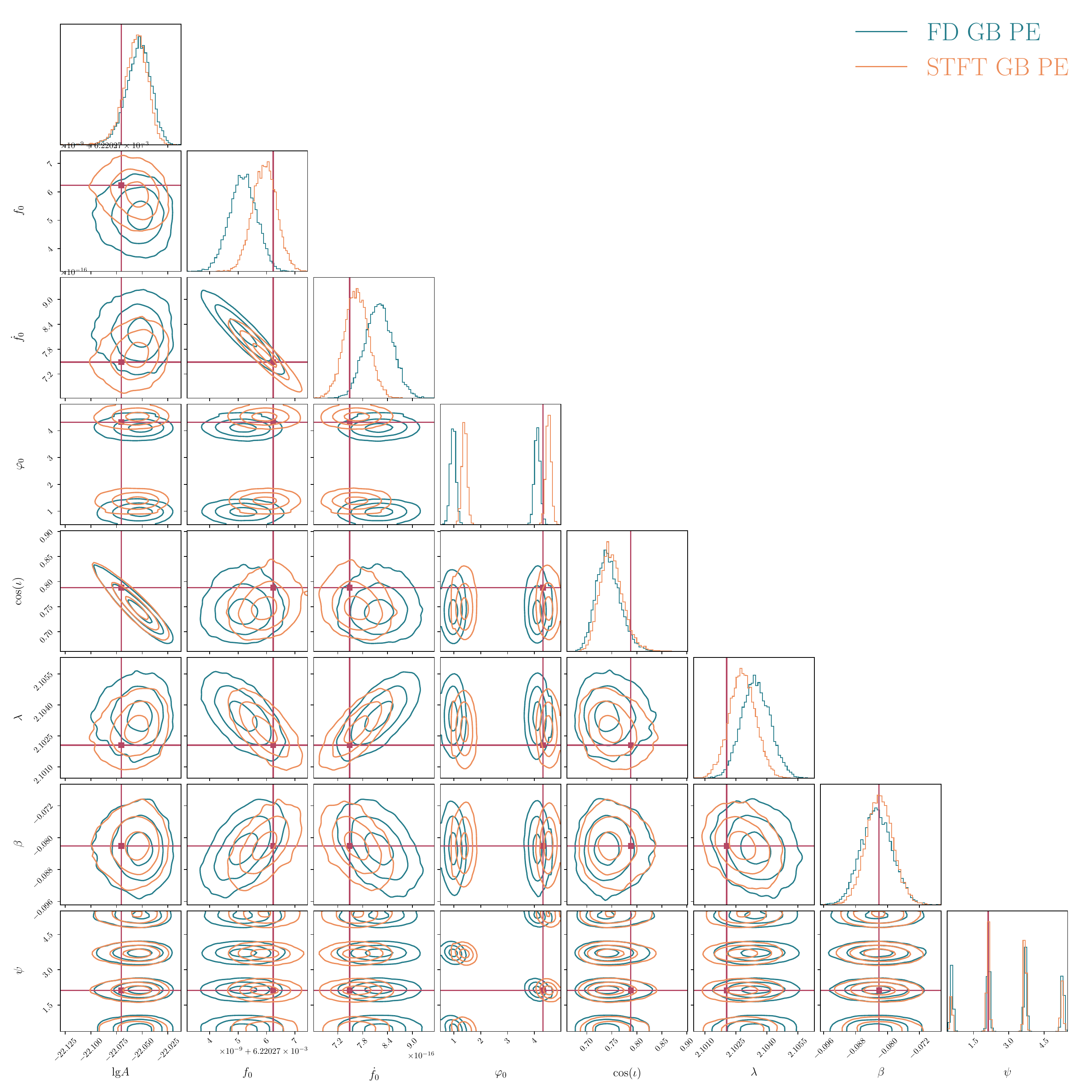}
    \caption{The posterior distributions of high-frequency source HMCnc ($f_0 = $ 6.22 mHz) drawn from frequency-domain (green) and STFT (orange) likelihoods, respectively.}
    \label{fig:HMCnc}
\end{figure}

\begin{figure}[h]
    \centering
    \includegraphics[width=0.8\linewidth]{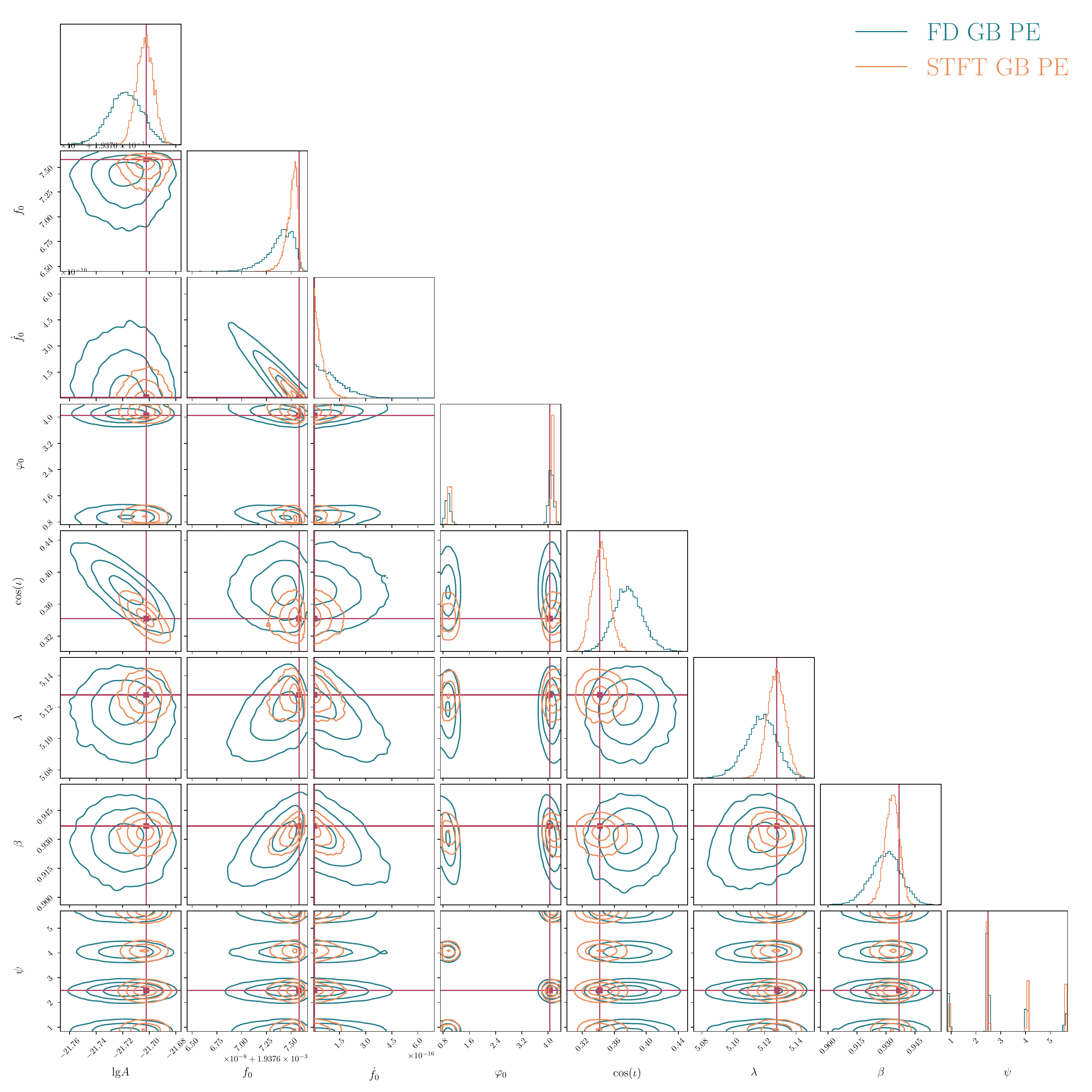}
    \caption{Similar to FIG.~\ref{fig:HMCnc}, but for a mid-low-frequency source ZTF J1905 ($f_0 = $ 1.94 mHz).}
    \label{fig:ZTFJ1905}
\end{figure}

\end{widetext}

\end{document}